\documentclass[twocolumn,superscriptaddress,preprintnumbers,nofootinbib]{revtex4}
\usepackage{amsmath,amssymb,longtable}
\usepackage{graphicx}
\usepackage{epsfig,color}
\usepackage{ulem}
\usepackage{bm}
\usepackage{longtable}
\usepackage{verbatim}

\begin{document}

\title{Quark structure of the $X(4500)$, $X(4700)$ and $\chi_{\rm c}(4P,5P)$ states}

\author{J. Ferretti}\email[]{jferrett@jyu.fi}
\affiliation{Department of Physics, University of Jyv\"askyl\"a, P.O. Box 35 (YFL), 40014 Jyv\"askyl\"a, Finland}

\author{E. Santopinto}\email[]{santopinto@ge.infn.it}
\affiliation{INFN, Sezione di Genova, Via Dodecaneso 33, 16146 Genova, Italy}

\begin{abstract}
We study some of the main properties (masses and open-flavor strong decay widths) of $4P$ and $5P$ charmonia. While there are two candidates for the $\chi_{\rm c0}(4P,5P)$ states, the $X(4500)$ and $X(4700)$, the properties of the other members of the $\chi_{\rm c}(4P,5P)$ multiplets are still completely unknown. With this in mind, we start to explore the charmonium interpretation for these mesons. Our second goal is to investigate if the apparent mismatch between the Quark Model (QM) predictions for $\chi_{\rm c0}(4P,5P)$ states and the properties of the $X(4500)$ and $X(4700)$ mesons can be overcome by introducing threshold corrections in the QM formalism.
According to our coupled-channel model results for the threshold mass shifts, the $\chi_{\rm c0}(5P) \rightarrow X(4700)$ assignment is unacceptable, while the $\chi_{\rm c0}(4P) \rightarrow X(4500)$ or $X(4700)$ assignments cannot be completely ruled out.
\end{abstract}

\maketitle

\section{Introduction}
In the past few years, our knowledge of the heavy-light and fully-heavy baryon and meson spectra has considerably improved \cite{Zyla:2020zbs}.
A large fraction of the newly-discovered hadrons perfectly fits into a standard quark-antiquark or three valence quark  description.
Some examples include the recently discovered $\Omega_{\rm c}$s \cite{Aaij:2017nav,Yelton:2017qxg}, the $\Xi_{\rm b}(6227)^-$ \cite{Aaij:2018yqz} and $\Sigma_{\rm b}(6097)^\pm$ \cite{Aaij:2018tnn} baryon states, and the $\chi_{\rm b1,2}(3P)$ heavy quarkonium resonances \cite{Aad:2011ih,Abazov:2012gh,Sirunyan:2018dff}.
There are, however, strong indications of the existence of exotic hadron configurations, which cannot be interpreted in terms of conventional quark-antiquark or three-quark degrees of freedom.
They include tetraquark and pentaquark candidates \cite{Zyla:2020zbs,Esposito:2016noz,Ali:2017jda,Olsen:2017bmm,Guo:2017jvc,Liu:2019zoy}, and suspected hybrid and glue-ball states \cite{Zyla:2020zbs,Esposito:2016noz,Ali:2017jda,Olsen:2017bmm,Liu:2019zoy}.

An important fraction of the suspected exotic mesons, the so-called $XYZ$ states, may require the introduction of complicated multiquark structures.
The most famous example is the $X(3872)$ [now $\chi_{\rm c1}(3872)$] \cite{Choi:2003ue,Acosta:2003zx,Abazov:2004kp}, but one could also mention the $X(4274)$ [also known as $\chi_{\rm c1}(4274)$] \cite{Aaij:2016iza,Aaltonen:2011at}.
Some of these exotics, the $Z_{\rm b}$ and $Z_{\rm c}$ resonances, like the $Z_c(3900)$ \cite{Ablikim,Liu}, $Z_b(10610)$ and $Z_b(10650)$ \cite{Bondar}, are characterized by very peculiar quark structures. $Z_{\rm Q}$ exotics are charged particles and, because of their energy and decay properties, they must contain a heavy $Q \bar Q$ pair (with $Q = c$ or $b$) too; thus, their adequate description requires the introduction of $Q \bar Q q\bar q$ four-quark configurations, where $q$ are light ($u$ or $d$) quarks.
If $Z_{\rm b}$ and $Z_{\rm c}$ states exist, one may also expect the emergence of hidden-charm/bottom tetraquarks with non-null strangeness content, the so-called $Z_{\rm cs}$ and $Z_{\rm bs}$ mesons; for example, see Refs. \cite{Voloshin:2019ilw,Ferretti:2020ewe}.
Recent indications of the possible existence of $Z_{\rm cs}$ states have been given by BESIII Collaboration \cite{Ablikim:2020hsk}.

In this paper, we study the main properties (masses, open-flavor and radiative decay widths) of the $4P$ and $5P$ charmonium multiplets.
While there are two candidates for the $\chi_{\rm c0}(4P,5P)$ states, the $X(4500)$ and $X(4700)$ resonances [also known as $\chi_{\rm c0}(4500)$ and $\chi_{\rm c0}(4700)$] \cite{Zyla:2020zbs,Aaij:2016iza,Aaij:2016nsc}, the properties of the other members of the $\chi_{\rm c}(4P,5P)$ multiplets are still completely unknown.
With this in mind, we start to explore the quark-antiquark interpretation for these mesons by computing their open-flavor strong decay widths. 
Our predictions may help the experimentalists in their search for the still unobserved $\chi_{\rm c}(4P,5P)$ resonances.
The calculation of the $\chi_{\rm c0}(4P,5P)$ radiative and hidden-flavor decay widths will be the subject of a subsequent paper.

We also provide Coupled-Channel Model (CCM) \cite{Ferretti:2018tco,Ferretti:2020civ} predictions for the physical masses\footnote{The physical masses of heavy quarkonia are the sum of a bare energy term and a self-energy/threshold correction.} of $4P$ and $5P$ charmonia, which may serve as a test for the $\chi_{\rm c0}(4P,5P) \rightarrow X(4500)$ or $X(4700)$ controversial assignments.
According to our CCM results, the introduction of threshold effects can hardly reconcile the Relativized Quark Model (RQM) predictions for the $\chi_{\rm c0}(4P,5P)$ meson masses \cite{Godfrey:1985xj} with the properties of the experimentally observed $X(4500)$ and $X(4700)$ states \cite{Zyla:2020zbs,Aaij:2016iza,Aaij:2016nsc}.
Therefore, the two previous resonances are unlikely to be associated with $\chi_{\rm c0}(4P,5P)$ charmonia, with the possible exception of $\chi_{\rm c0}(4P) \rightarrow X(4500)$ or $X(4700)$.

There are several alternative interpretations for the $\chi_{\rm c0}(4500)$ and $\chi_{\rm c0}(4700)$ states \cite{Ortega:2016hde,Lu:2016cwr,Anwar:2018sol,Maiani:2016wlq,Chen:2016oma,Oncala:2017hop,Liu:2016onn}.
A possible explanation of the $\chi_{\rm c0}(4500)$ and $\chi_{\rm c0}(4700)$ unusual properties without resorting to exotic interpretations may be to hypothesize a progressive departure of the $c \bar c$ linear confining potential from the $\propto r$ behavior as one goes up in energy. This departure could be either due to limitations of the relativized QM fit \cite{Godfrey:1985xj}, which little by little make their appearance at higher meson energies, or to the need of renormalizing the $c \bar c$ color string tension at higher energies to take relativistic effects (like $q \bar q$ light quark pair creation) explicitly into account. For example, see Ref. \cite{Geiger:1989yc}.

The $X(4500)$ and $X(4700)$ were interpreted as compact tetraquarks in Refs. \cite{Lu:2016cwr,Anwar:2018sol,Maiani:2016wlq,Chen:2016oma}.
In particular, in Ref. \cite{Anwar:2018sol} the authors made use of a relativized diquark model to calculate the spectrum of hidden-charm tetraquarks. According to their findings, the $X(4500)$ and $X(4700)$ can be described as $0^{++}$ radial excitations of $S$-wave axial-vector diquark-antidiquark and scalar diquark-antidiquark bound states, respectively.
A similar interpretation was provided in Ref. \cite{Lu:2016cwr}.
Stancu calculated the $s\bar s c \bar c$ tetraquark spectrum within a quark model with chromomagnetic interaction \cite{Stancu:2009ka}.
She interpreted the $X(4140)$ as the strange partner of the $X(3872)$, but she could not accommodate the other $s\bar s c \bar c$ states, the $X(4274)$, $X(4500)$ and $X(4700)$.\footnote{The $X(4500)$ and $X(4700)$ were observed at LHCb in $2016$~\cite{Aaij:2016iza}, and the $X(4274)$ was first observed in $2011$ by CDF with a small significance of $3.1\sigma$~\cite{Aaltonen:2011at}, while Stancu's analysis dates back to 2010.}
By using QCD sum rules, the $X(4500)$ and $X(4700)$ were interpreted as $D$-wave $c\bar c s\bar s$ tetraquark states with opposite color structures~\cite{Chen:2016oma}.
Maiani {\it et al.} could accommodate the $X(4140)$, $X(4274)$, $X(4500)$ and $X(4700)$ in two tetraquark multiplets.
They also suggested that the $X(4500)$ and $X(4700)$ are $2S$ $cs \bar c \bar s$ tetraquark states \cite{Maiani:2016wlq}.

In Ref.~\cite{Ortega:2016hde}, the authors investigated possible assignments for the four $J/\psi \phi$ structures reported by LHCb \cite{Aaij:2012pz} in a coupled channel scheme by using a nonrelativistic constituent quark model \cite{Vijande:2004he}.\footnote{Four $J/\psi \phi$ structures were reported by LHCb only on the basis of a 6D amplitude analysis \cite{Aaij:2012pz}. A narrow $X(4140)$ was reported by CDF \cite{Aaltonen:2009tz} and then confirmed by D0 \cite{Abazov:2013xda}. BaBar did not see anything statistically significant \cite{Lees:2014lra}. CMS confirmed a slightly broader X(4140) and a less significant second peak \cite{Chatrchyan:2013dma}. The LHCb amplitude analysis supersedes all this, and finds a much broader X(4140) \cite{Aaij:2016iza}.} 
In particular, they showed that the $X(4274)$, $X(4500)$ and $X(4700)$ mesons can be described as conventional $3^3P_1$, $4^3P_0$, and $5^3P_0$ charmonium states, respectively.
In Ref. \cite{Liu:2016onn}, the author studied the nature of the $X(4140)$, $X(4274)$, $X(4500)$, and $X(4700)$ states in the process $B^+ \rightarrow J/\psi \phi K^+$ by means of the rescattering mechanism. 
According to his results, the properties of the $X(4700)$ and $X(4140)$ can be explained by the rescattering effects, while those of the $X(4274)$ and $X(4500)$ cannot if the quantum numbers of the $X(4274)$ and $X(4500)$ are $1^{++}$ and $0^{++}$, respectively. This indicates that, unlike the $X(4700)$ and $X(4140)$, the $X(4274)$ and $X(4500)$ could be genuine resonances.

In the study of heavy quarkonium hybrids based on the strong coupling regime of potential nonrelativistic QCD of Ref. \cite{Oncala:2017hop}, the authors found that most of the isospin zero $XYZ$ states fit well either as the hybrid or standard quarkonium candidates.
According to their results, the $X(4500)$ is compatible with a $0^{++}$ hybrid state, even though its mixing with the spin-1 charmonium is little and it is difficult to understand its observation in the $J/\psi \phi$ channel; the $X(4700)$ is compatible with the charmonium $\chi_{\rm c0}(4P)$.

Finally, it is worth to remind that both the $X(4500)$ and $X(4700)$ are omitted from the PDG summary table \cite{Zyla:2020zbs}.
This means that their existence still needs to be proved. Future experimental searches may thus confirm their presence at similar or slightly different energies or even rule out their existence.

\section{Open-flavor strong decays of $4P$ and $5P$ charmonium states}
\label{3P0-formulas}
Our analysis starts with the calculation of the open-charm strong decays of the $\chi_{\rm c}(4P,5P)$ states within the $^3P_0$ pair-creation model \cite{Micu,LeYaouanc,Roberts:1992}.
Open-charm are usually the dominant decay modes of hadron higher radial excitations; the contributions of hidden-charm and radiative decay modes to the total width of a higher-lying charmonium state are indeed expected to be in the order of a few percent or even less.
This is why the calculated open-flavor total decay widths of higher charmonia are precious informations, which can be directly used for a comparison with the experimental total widths of those states within a reasonable grade of accuracy.

\begin{figure}[htbp]
\begin{center}
\includegraphics[width=8cm]{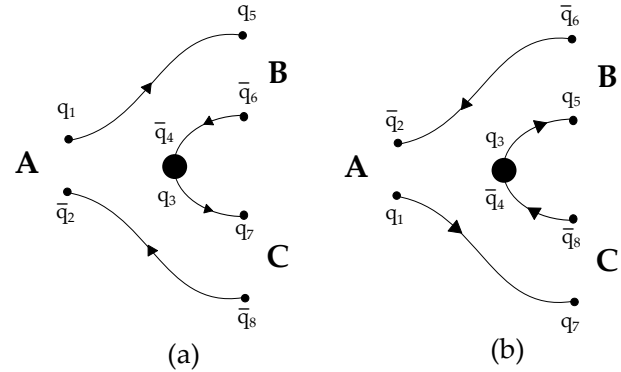}
\end{center}
\caption{Diagrams contributing to the $A\rightarrow BC$ open-flavor decay process. $q_{\rm i}$, with i=1, ..., 4, and $\bar{q}_{\rm j}$, with j=5, ..., 8, are the quarks and antiquarks in the parent and daughter states, respectively. Picture from Ref.~\cite{bottomonium}. APS copyright.} 
\label{fig:diagrammi3P0}
\end{figure}

In the $^3P_0$ pair-creation model, the open-flavor strong decay $A \rightarrow B C$ takes place in the rest frame of the parent hadron $A$ and proceeds via the creation of an additional $q \bar q$ pair (with $q = u, d$ or $s$) characterized by $J^{PC} = 0^{++}$ quantum numbers \cite{Micu,LeYaouanc,Roberts:1992} (see Fig. \ref{fig:diagrammi3P0}). 
The width is calculated as \cite{Micu,LeYaouanc,Ackleh:1996yt} 
\begin{equation}
	\label{eqn:3P0-decays-ABC}
	\Gamma_{A \rightarrow BC} = \Phi_{A \rightarrow BC}(k_0) \sum_{\ell} 
	\left| \left\langle BC k_0  \, \ell J \right| T^\dag \left| A \right\rangle \right|^2 \mbox{ },
\end{equation}
where $\ell$ is the relative angular momentum between the hadrons $B$ and $C$, ${\bf J} = {\bf J}_B + {\bf J}_C + {\bm\ell}$ is their total angular momentum, and
\begin{equation}
	\label{eqn:rel-PSF}
	\Phi_{A \rightarrow BC}(k_0) = 2 \pi k_0 \frac{E_B(k_0) E_C(k_0)}{M_A}  
\end{equation}
is the phase-space factor for the decay.
\begin{table}[htbp]  
\begin{center}
\begin{ruledtabular}
\begin{tabular}{ll} 
Parameter  & Value \\ 
\hline
$\gamma_0$ & 0.510 \\  
$\alpha_{\rm ho}$   & 0.500 GeV \\  
$\alpha_{\rm d}$   & 0.589 GeV  \\
$m_{\rm u,d}$ & 0.330 GeV \\
$m_{\rm s}$ & 0.550 GeV \\
$m_{\rm c}$ & 1.50 GeV    \\ 
\end{tabular}
\end{ruledtabular}
\end{center}
\caption{$^3P_0$ pair-creation model parameters for the charmonium sector, extracted from \cite[Table II]{Ferretti:2013faa} and \cite[Table II]{Ferretti:2015rsa}. The valence quark mass parameters, $m_{\rm i}$ (with i = u, d, s, c), are used in the calculation of the $\left\langle BC k_0  \, \ell J \right| T^\dag \left| A \right\rangle$ amplitudes of Eq. (\ref{eqn:3P0-decays-ABC}) and also appear in the expression of the effective pair-creation strength of Eq. (\ref{eqn:gamma0-eff}).}
\label{tab:3P0parameters}  
\end{table} 
Here, $k_0$ is the relative momentum between $B$ and $C$, $M_A$ and $E_{B,C}(k_0)$ are the energies of the parent and daughter hadrons, respectively.
We assume harmonic oscillator wave functions for the parent and daughter hadrons, $A$, $B$ and $C$, depending on a single oscillator parameter $\alpha_{\rm ho}$. The values of the oscillator parameter, $\alpha_{\rm ho}$, and of the other pair-creation model parameters, $\alpha_{\rm d}$ and $\gamma_0$, were fitted to the open-charm strong decays of higher charmonia \cite{Ferretti:2013faa} and also used later in the study of charmed and charmed-strange meson open-flavor strong decays \cite{Ferretti:2015rsa} and of the quasi two-body decay of the $X(3872)$ into $D^0 (\bar D^0 \bar \pi^0)_{\bar D^{*0}}$ \cite{Ferretti:2014xqa}; see Table \ref{tab:3P0parameters}. 

\begin{table}[htbp]  
\begin{center}
\begin{ruledtabular}
\begin{tabular}{cc} 
State  & Bare Mass [MeV] \\ 
\hline
$h_{\rm c}(4P)$ & 4634 \\  
$\chi_{\rm c0}(4P)$ & 4613 \\  
$\chi_{\rm c1}(4P)$ & 4633  \\
$\chi_{\rm c2}(4P)$ & 4650 \\
$h_{\rm c}(5P)$ & 4919 \\
$\chi_{\rm c0}(5P)$ & 4902 \\
$\chi_{\rm c1}(5P)$ & 4919 \\
$\chi_{\rm c2}(5P)$ & 4934 \\
\end{tabular}
\end{ruledtabular}
\end{center}
\caption{Bare masses of $\chi_{\rm c}(4P,5P)$ charmonia, computed in a variational program by using the original relativized QM parameters \cite[Table II]{Godfrey:1985xj}.}
\label{tab:bare-masses}  
\end{table}

\begin{table*}
\footnotesize
\centering
\begin{ruledtabular}
\begin{tabular}{cc|cc|cc|cc} 
$h_{\rm c}(4P)$ decay channel & Width & $\chi_{\rm c0}(4P)$ decay channel & Width & $\chi_{\rm c1}(4P)$ decay channel & Width & $\chi_{\rm c2}(4P)$ decay channel & Width \\
                                &           [MeV]                            &     &      [MeV]               &                                             &           [MeV] &                           &           [MeV]  \\                                
\hline
$D \bar D^*$ & 0.6 & $D \bar D$ & 0.8 & $D \bar D^*$ & 1.1 & $D \bar D$ & 2.7   \\ 
$D^* \bar D^*$ & 8.1 & $D^* \bar D^*$ & 13.1 & $D^* \bar D^*$ & 6.0 & $D \bar D^*$ & 0.5    \\   
$D \bar D_1(2^3S_1)$ & 28.9 & $D \bar D_0(2550)$ & 23.7 & $D \bar D_1(2^3S_1)$ & 18.9 & $D^* \bar D^*$ & 9.6    \\                                                                                                                                       
$D^* \bar D_0(2550)$ & 18.3 & $D \bar D_1(2420)$ & 25.5 & $D^* \bar D_0(2550)$ & 12.8 & $D \bar D_0(2550)$ & 4.6   \\ 
$D \bar D_0^*(2300)$ & 1.9$^\dag$ & $D \bar D_1(2430)$ & 9.0 & $D \bar D_0^*(2300)$ & 0.003$^\dag$ & $D \bar D_1(2^3S_1)$ & 22.2   \\       
$D \bar D_1(2420)$ & 0.01$^\dag$ & $D^* \bar D_0^*(2300)$ & 4.3 & $D \bar D_1(2420)$ & 3.3 & $D^* \bar D_0(2550)$ & 19.2     \\                                                                                                                                                  
$D \bar D_1(2430)$ & 0.05$^\dag$ & $D^* \bar D_1(2420)$ & 0.4$^\dag$ & $D \bar D_1(2430)$ & 8.2 & $D \bar D_1(2420)$ & 4.5    \\                                                                                                                                                                  
$D \bar D_2^*(2460)$ & 24.7 & $D^* \bar D_1(2430)$ & 0.2$^\dag$ & $D \bar D_2^*(2460)$ & 19.6 & $D \bar D_1(2430)$ & 3.3    \\                                                                                                                                          
$D^* \bar D_0^*(2300)$ & 0.04$^\dag$ & $D^* \bar D_2^*(2460)$ & 72.6 & $D^* \bar D_0^*(2300)$ & 4.0 & $D \bar D_2^*(2460)$ & 9.1    \\                                                                                                                                                                                                                                                                                                                                                                           
$D^* \bar D_1(2420)$ & 23.5 & $D_0^*(2300) \bar D_0^*(2300)$ & 0.6$^\dag$ & $D^* \bar D_1(2420)$ & 13.8 & $D^* \bar D_0^*$ & 3.1   \\                                                                     
$D^* \bar D_1(2430)$ & 15.4 & $D_{\rm s} \bar D_{\rm s}$ & 1.1 & $D^* \bar D_1(2430)$ & 8.7 & $D^* \bar D_1(2420)$ & 19.8    \\                                       
$D^* \bar D_2^*(2460)$ & 31.7 & $D_{\rm s}^* \bar D_{\rm s}^*$ & 1.7 & $D^* \bar D_2^*(2460)$ & 57.8 & $D^* \bar D_1(2430)$ & 13.4    \\                                                                                                                                                                                                                                                                                                     
$D \bar D_3(2750)$ & 0.02$^\dag$ & $D_{\rm s} \bar D_{\rm s1}(2460)$ & 0.05 & $D \bar D_3(2750)$ & 0.005$^\dag$ & $D^* \bar D_2^*(2460)$ & 27.2 \\                                                                                                                                      
$D_{\rm s} \bar D_{\rm s}^*$ & 2.6 & $D_{\rm s} \bar D_{\rm s1}(2536)$ & 0.6 & $D_{\rm s} \bar D_{\rm s}^*$ & 2.5 & $D_0^*(2300) \bar D_0^*(2300)$ & 0.002$^\dag$   \\                                   
$D_{\rm s}^* \bar D_{\rm s}^*$ & 1.3 & $D_{\rm s}^* \bar D_{\rm s0}^*(2317)$ & 0.02 & $D_{\rm s}^* \bar D_{\rm s}^*$ & 1.6 & $D \bar D_3(2750)$ & 0.4$^\dag$    \\                                                                    
$D_{\rm s} \bar D_{\rm s0}^*(2317)$ & 0.6 & $D_{\rm s}^* \bar D_{\rm s1}(2460)$ & 1.1 & 
$D_{\rm s} \bar D_{\rm s0}^*(2317)$ & 0.02 & $D_{\rm s} \bar D_{\rm s}$ & 0.1    \\  
$D_{\rm s} \bar D_{\rm s1}(2536)$ & 0.01 & & & $D_{\rm s} \bar D_{\rm s1}(2460)$ & 0.03 & $D_{\rm s} \bar D_{\rm s}^*$ & 1.6      \\   
$D_{\rm s} \bar D_{\rm s2}^*(2573)$ & 4.3 & & & $D_{\rm s} \bar D_{\rm s1}(2536)$ & 1.1 & $D_{\rm s}^* \bar D_{\rm s}^*$ & 1.5  \\                                                                                                                                                                                                                                                                                                                 
$D_{\rm s}^* \bar D_{\rm s0}^*(2317)$ & 0.002 & & & $D_{\rm s} \bar D_{\rm s2}^*(2573)$ & 3.8 & $D_{\rm s} \bar D_{\rm s0}(2^1S_0)$ & 0.01  \\
$D_{\rm s}^* \bar D_{\rm s1}(2460)$ & 6.0 & & & $D_{\rm s}^* \bar D_{\rm s0}^*(2317)$ & 0.006 & $D_{\rm s} \bar D_{\rm s1}(2460)$ & 2.7\\    
& &  &  & $D_{\rm s}^* \bar D_{\rm s1}(2460)$ & 7.8 & $D_{\rm s} \bar D_{\rm s1}(2536)$ & 1.1 \\  
& &  &  & &  & $D_{\rm s} \bar D_{\rm s2}^*(2573)$ & 1.8 \\  
 &  &                              &  &                            &     & $D_{\rm s}^* \bar D_{\rm s0}^*(2317)$ & 0.02 \\                                                                  
 &  &                              &  &                   &              & $D_{\rm s}^* \bar D_{\rm s1}(2460)$ & 2.5    \\  
 &  &                              &  &                   &              & $D_{\rm s}^* \bar D_{\rm s1}(2536)$ & 0.09    \\  
 &  &                              &  &                            &     & $D_{\rm s0}^*(2317) \bar D_{\rm s0}^*(2317)$ & $5\cdot10^{-8}$ \\ 
\hline                                 
tot open-flavor & 168 & tot open-flavor & 155 & tot open-flavor & 171 & tot open-flavor & 151 \\                                                                                                                                                                                                                                                                                                                                                                                                                                                                                                                                                                                                            
\end{tabular}
\end{ruledtabular}
\caption{Open-charm strong decays of $\chi_{\rm c}(4P)$ states in the $^3P_0$ pair-creation model.  The values of the $\chi_{\rm c}(4P)$ masses are calculated in the relativized QM of Ref.~\cite{Godfrey:1985xj} and are reported in Table \ref{tab:bare-masses}. 
The values of the charmed and charmed-strange meson masses are taken from the PDG \cite{Zyla:2020zbs} (when available) or from the relativized QM calculation of \cite[Tables III and IV]{Ferretti:2015rsa}. The entries marked by the symbol $^\dag$ may not be reliable; the calculation of these decay widths may require averaging over the Breit-Wigner distributions of the daughter mesons. For more details, see the discussion below.}
\label{tab:ChiC(4P)-strong}  
\end{table*}

Some changes are introduced in the original form of the $^3P_0$ pair-creation model operator, $T^\dag$.
They include: I) the substitution of the pair-creation strength, $\gamma_0$, with an effective one \cite{bottomonium}, $\gamma_0^{\rm eff}$, to suppress heavy quark pair-creation \cite{bottomonium,Kalashnikova:2005ui,Strong2015}; II) the introduction of a Gaussian quark form-factor, because the $q \bar q$ pair of created quarks has an effective size \cite{bottomonium,Geiger:1989yc,Bijker:2009up,Strong2015}. More details on the $^3P_0$ pair-creation model can be found in Appendix \ref{3P0model}.

When available, we extract the masses of the parent and daughter mesons from the PDG \cite{Zyla:2020zbs}; otherwise, we calculate them by using the relativized QM with the original values of its parameters; see \cite[Table II]{Godfrey:1985xj}.
The masses of the $\chi_{\rm c0}(4500)$ and $\chi_{\rm c0}(4700)$ resonances \cite{Zyla:2020zbs,Aaij:2016iza,Aaij:2016nsc}, $4506\pm11^{+12}_{-15}$ MeV and $4704\pm10^{+14}_{-24}$ MeV, seem to be incompatible with the relativized QM predictions for the $\chi_{\rm c0}(4P,5P)$ states; see Table \ref{tab:bare-masses}.
A coupled-channel model calculation, with the goal of reconciling relativized QM predictions and the experimental data, is carried out in Sec. \ref{Threshold mass-shifts}.

Given the previous apparent incompatibility, in the $\chi_{\rm c0}(4P,5P)$ cases we provide results by using: I) the relativized QM values of the masses from Table \ref{tab:bare-masses}; II) the tentative assignments $\chi_{\rm c0}(4500) \rightarrow \chi_{\rm c0}(4P)$ and $\chi_{\rm c0}(4700) \rightarrow \chi_{\rm c0}(4P)$ or $\chi_{\rm c0}(5P)$, with the experimental values of the $\chi_{\rm c0}(4500)$ and $\chi_{\rm c0}(4700)$ masses as inputs in the calculation.

The mixing angles between $1^1P_1$ and $1^3P_1$, $2^1P_1$ and $2^3P_1$ and also $2^1D_2$ and $2^3D_2$ charmed and charmed-strange states are taken from \cite[Tables III and IV]{Ferretti:2015rsa}. 
In the case of $1P$ charmed-strange mesons, the mass difference between $\left |1P_1 \right\rangle$ and $\left|1P_1' \right\rangle$ states (75 MeV) is much larger than that in the charmed sector (6 MeV). Thus, for $1P$ charmed-strange mesons we make use of the approximation: $\left |1P_1 \right\rangle \simeq \left |1^1P_1\right\rangle$ and $\left|1P_1' \right\rangle \simeq \left|1^3P_1 \right\rangle$.
\begin{table*}
\footnotesize
\centering
\begin{ruledtabular}
\begin{tabular}{cc|cc|cc|cc} 
$h_{\rm c}(5P)$ decay channel & Width & $\chi_{\rm c0}(5P)$ decay channel & Width & $\chi_{\rm c1}(5P)$ decay channel & Width & $\chi_{\rm c2}(5P)$ decay channel & Width \\
\hline                                     
$D \bar D^*$ & 1.2 & $D \bar D$ & 0.01 & $D \bar D^*$ & 1.9 & $D \bar D$ & 0.4   \\ 
$D^* \bar D^*$ & 5.7 & $D^* \bar D^*$ & 8.8 & $D^* \bar D^*$ & 5.0 & $D \bar D^*$ & 0.2    \\    
$D^* \bar D_0(2550)$ & 10.5 & $D \bar D_0(2550)$ & 9.9 & $D^* \bar D_0(2550)$ & 8.6 & $D^* \bar D^*$ & 6.8    \\        
$D \bar D_1(2^3S_1)$ & 11.6 & $D^* \bar D_1(2^3S_1)$ & 4.5 & $D \bar D_1(2^3S_1)$ & 11.2 & $D \bar D_0(2550)$ & 1.5$^\dag$  \\   
$D^* \bar D_1(2^3S_1)$ & 4.9 & $D \bar D_1(2420)$ & 12.7 & $D^* \bar D_1(2^3S_1)$ & 7.3 & $D^* \bar D_0(2550)$ & 7.8   \\ 
$D \bar D_0^*(2300)$ & 1.6 & $D \bar D_1(2430)$ & 4.0 & $D \bar D_0^*(2300)$ & 0.002$^\dag$ & $D \bar D_1(2^3S_1)$ & 6.7     \\   
$D \bar D_1(2420)$ & 0.004$^\dag$ & $D^* \bar D_0^*(2300)$ & 2.7 & $D \bar D_1(2420)$ & 1.5 & $D^* \bar D_1(2^3S_1)$ & 6.8     \\                                                                                    
$D \bar D_1(2430)$ & 0.02$^\dag$ & $D^* \bar D_1(2420)$ & 1.2 & $D \bar D_1(2430)$ & 4.4 & $D \bar D_1(2420)$ & 3.5 \\                                                                                                                                                                  
$D \bar D_2^*(2460)$ & 12.6 & $D^* \bar D_1(2430)$ & 2.5 & $D \bar D_2^*(2460)$ & 9.7 & $D \bar D_1(2430)$ & 2.6    \\                                                                                                                                          
$D^* \bar D_0^*(2300)$ & 0.01$^\dag$ & $D^* \bar D_2^*(2460)$ & 21.9 & $D^* \bar D_0^*(2300)$ & 2.5 & $D \bar D_2^*(2460)$ & 5.1 \\                                                                                                                                                                                                                                                                                                                                                                           
$D^* \bar D_1(2420)$ & 10.8 & $D \bar D_1(2P_1)$ & 5.6 & $D^* \bar D_1(2420)$ & 8.4 & $D^* \bar D_0^*(2300)$ & 2.1   \\                                                                     
$D^* \bar D_1(2430)$ & 7.7 & $D \bar D_1(2P_1')$ & 4.4 & $D^* \bar D_1(2430)$ & 5.0 & $D^* \bar D_1(2420)$ & 8.2  \\                                       
$D^* \bar D_2^*(2460)$ & 11.6 & $D_0^*(2300) \bar D_0^*(2300)$ & 0.002$^\dag$ & $D^* \bar D_2^*(2460)$ & 18.2 & $D^* \bar D_1(2430)$ & 6.0 \\                                                                                                                                                                                                                                                                                                     
$D_0(2550) \bar D_0^*(2300)$ & 2.6 & $D_0^*(2300) \bar D_2^*(2460)$ & 2.9 & $D_0(2550) \bar D_0^*(2300)$ & 0.06$^\dag$ & $D^* \bar D_2^*(2460)$ & 14.8 \\       
$D \bar D_0(2^3P_0)$ & 0.4 & $D_1(2420) \bar D_1(2420)$ & 11.0 & $D \bar D_0(2^3P_0)$ & 0.001 & $D \bar D_1(2P_1)$ & 11.9  \\    
$D \bar D_1(2P_1)$ & 0.008 & $D_1(2420) \bar D_1(2430)$ & 7.5 & $D \bar D_1(2P_1)$ & 1.5 & $D \bar D_1(2P_1')$ & 7.7    \\  
$D \bar D_1(2P_1')$ & 0.02 & $D_1(2430) \bar D_1(2430)$ & 5.1 & $D \bar D_1(2P_1')$ & 4.7 & $D \bar D_2(2^3P_2)$ & 11.5    \\   
$D \bar D_2(2^3P_2)$ & 20.9 & $D_1(2420) \bar D_2^*(2460)$ & 0.4$^\dag$ & $D \bar D_2(2^3P_2)$ & 14.1 & $D^* \bar D_1(2P_1)$ & 0.1   \\   
$D_0^*(2300) \bar D_1(2420)$ & 1.0$^\dag$ & $D_1(2430) \bar D_2^*(2460)$ & 0.5$^\dag$ & $D_0^*(2300) \bar D_1(2420)$ & 0.2$^\dag$ & $D_0^*(2300) \bar D_0^*(2300)$ & 0.02$^\dag$    \\   
$D_0^*(2300) \bar D_1(2430)$ & 0.3$^\dag$ & $D \bar D_2(1D_2)$ & 0.2 & $D_0^*(2300) \bar D_1(2430)$ & 0.7$^\dag$ & $D_0^*(2300) \bar D_1(2420)$ & 0.02$^\dag$   \\ 
$D_0^*(2300) \bar D_2^*(2460)$ & 0.009$^\dag$ & $D \bar D_2(1D_2')$ & 0.3 & $D_0^*(2300) \bar D_2^*(2460)$ & 1.4$^\dag$ & $D_0^*(2300) \bar D_1(2430)$ & 0.07$^\dag$   \\ 
$D_1(2420) \bar D_1(2420)$ & 0.06$^\dag$ & $D^* \bar D_3(2750)$ & 17.3 & $D_1(2420) \bar D_1(2420)$ & 2.6 & $D_0^*(2300) \bar D_2^*(2460)$ & 1.5$^\dag$   \\ 
$D_1(2420) \bar D_1(2430)$ & 0.3$^\dag$ & $D^* \bar D_1(1^3D_1)$ & 0.4 & $D_1(2420) \bar D_1(2430)$ & 7.3  & $D_1(2420) \bar D_1(2420)$ & 2.7   \\ 
$D_1(2430) \bar D_1(2430)$ & 1.1$^\dag$ & $D^* \bar D_2(1D_2)$ & 6.5 & $D_1(2430) \bar D_1(2430)$ & 3.6 & $D_1(2420) \bar D_1(2430)$ & 3.0  \\ 
$D_1(2420) \bar D_2^*(2460)$ & 9.8 & $D^* \bar D_2(1D_2')$ & 7.2 & $D_1(2420) \bar D_2^*(2460)$ & 8.9 & $D_1(2430) \bar D_1(2430)$ & 1.8  \\ 
$D_1(2430) \bar D_2^*(2460)$ & 6.4 & $D_{\rm s} \bar D_{\rm s}$ & 0.5 & $D_1(2430) \bar D_2^*(2460)$ & 4.9 & $D_1(2420) \bar D_2^*(2460)$ & 5.3   \\ 
$D \bar D_3(2750)$ & 11.7 & $D_{\rm s}^* \bar D_{\rm s}^*$ & 0.3 & $D \bar D_3(2750)$ & 6.8 & $D_1(2430) \bar D_2^*(2460)$ & 4.1   \\ 
$D \bar D_1(1^3D_1)$ & 0.2 & $D_{\rm s} \bar D_{\rm s0}(2^1S_0)$ & 0.4 & $D \bar D_1(1^3D_1)$ & 0.08 & $D_2^*(2460) \bar D_2^*(2460)$ & 1.1   \\ 
$D \bar D_2(1D_2)$ & $6\cdot10^{-5}$ & $D_{\rm s}^* \bar D_{\rm s1}^*(2700)$ & 10.5 & $D \bar D_2(1D_2)$ & 0.2 & $D \bar D_3(2750)$ & 5.2    \\                                                                                                                                                                                             
$D \bar D_2(1D_2')$ & $9\cdot10^{-5}$ & $D_{\rm s} \bar D_{\rm s1}(2460)$ & 0.07 & $D \bar D_2(1D_2')$ & 0.2 & $D \bar D_1(1^3D_1)$ & $3\cdot10^{-8}$    \\ 
$D^* \bar D_3(2750)$ & 14.5 & $D_{\rm s} \bar D_{\rm s1}(2536)$ & 0.01 & $D^* \bar D_3(2750)$ & 16.9 & $D \bar D_2(1D_2)$ & 5.3    \\ 
$D^* \bar D_1(1^3D_1)$ & 0.5 & $D_{\rm s}^* \bar D_{\rm s0}^*(2317)$ & 0.01 & $D^* \bar D_1(1^3D_1)$ & 1.0 & $D \bar D_2(1D_2')$ & 4.9      \\ 
$D^* \bar D_2(1D_2)$ & 10.8 & $D_{\rm s}^* \bar D_{\rm s1}(2460)$ & 0.1 & $D^* \bar D_2(1D_2)$ & 12.2 & $D^* \bar D_3(2750)$ & 17.6    \\ 
$D^* \bar D_2(1D_2')$ & 9.4 & $D_{\rm s}^* \bar D_{\rm s1}(2536)$ & 1.8 & $D^* \bar D_2(1D_2')$ & 10.1 & $D^* \bar D_1(1^3D_1)$ & 1.3 \\ 
$D_{\rm s} \bar D_{\rm s}^*$ & 1.0 & $D_{\rm s}^* \bar D_{\rm s2}^*(2573)$ & 5.4 & $D_{\rm s} \bar D_{\rm s}^*$ & 0.9 & $D^* \bar D_2(1D_2)$ & 4.7 \\ 
$D_{\rm s}^* \bar D_{\rm s}^*$ & 0.3 & $D_{\rm s0}^*(2317) \bar D_{\rm s0}^*(2317)$ & 0.01 & $D_{\rm s}^* \bar D_{\rm s}^*$ & 0.4 & $D^* \bar D_2(1D_2')$ & 4.1 \\ 
$D_{\rm s} \bar D_{\rm s1}^*(2700)$ & 0.2 & $D_{\rm s0}^*(2317) \bar D_{\rm s2}^*(2573)$ & 0.02$^\dag$ & $D_{\rm s} \bar D_{\rm s1}^*(2700)$ & 0.4$^\dag$ & $D_{\rm s} \bar D_{\rm s}$ & 0.2 \\   
$D_{\rm s}^* \bar D_{\rm s0}(2^1S_0)$ & 2.9 & $D_{\rm s} \bar D_{\rm s2}(1D_2)$ & 0.5 & $D_{\rm s}^* \bar D_{\rm s0}(2^1S_0)$ & 2.5 & $D_{\rm s} \bar D_{\rm s}^*$ & 0.7  \\    
$D_{\rm s}^* \bar D_{\rm s1}^*(2700)$ & 6.0 & $D_{\rm s} \bar D_{\rm s2}(1D_2')$ & 0.3 & $D_{\rm s}^* \bar D_{\rm s1}^*(2700)$ & 8.2 & $D_{\rm s}^* \bar D_{\rm s}^*$ & 0.3  \\   
$D_{\rm s} \bar D_{\rm s0}^*(2317)$ & 0.3 & & & $D_{\rm s} \bar D_{\rm s0}^*(2317)$ & 0.006 & $D_{\rm s} \bar D_{\rm s0}(2^1S_0)$ & 0.2  \\    
$D_{\rm s} \bar D_{\rm s1}(2536)$ & $5\cdot10^{-5}$ & & & $D_{\rm s} \bar D_{\rm s1}(2460)$ & 0.03 & $D_{\rm s} \bar D_{\rm s1}^*(2700)$ & 0.02$^\dag$ \\ 
$D_{\rm s} \bar D_{\rm s2}^*(2573)$ & 0.3 & & & $D_{\rm s} \bar D_{\rm s1}(2536)$ & 0.07 & $D_{\rm s}^* \bar D_{\rm s0}(2^1S_0)$ & 1.6 \\ 
$D_{\rm s}^* \bar D_{\rm s0}^*(2317)$ & 0.001 & & & $D_{\rm s} \bar D_{\rm s2}^*(2573)$ & 0.3$^\dag$ & $D_{\rm s}^* \bar D_{\rm s1}^*(2700)$ & 6.2 \\ 
$D_{\rm s}^* \bar D_{\rm s1}(2460)$ & 0.7 & & & $D_{\rm s}^* \bar D_{\rm s0}^*(2317)$ & 0.03 & $D_{\rm s} \bar D_{\rm s1}(2460)$ & 0.7  \\     
$D_{\rm s}^* \bar D_{\rm s1}(2536)$ & 1.5 & & & $D_{\rm s}^* \bar D_{\rm s1}(2460)$ & 0.9 & $D_{\rm s} \bar D_{\rm s1}(2536)$ & 0.1 \\
$D_{\rm s}^* \bar D_{\rm s2}^*(2573)$ & 2.9 & & & $D_{\rm s}^* \bar D_{\rm s1}(2536)$ & 1.0 & $D_{\rm s} \bar D_{\rm s2}^*(2573)$ & 0.1$^\dag$ \\                                                                                                                                                                                                                                                                                                            
$D_{\rm s0}^*(2317) \bar D_{\rm s1}(2460)$ & 0.5 & & & $D_{\rm s}^* \bar D_{\rm s2}^*(2573)$ & 3.2 & $D_{\rm s}^* \bar D_{\rm s0}^*(2317)$ & 0.03     \\
$D_{\rm s0}^*(2317) \bar D_{\rm s1}(2536)$ & 0.02 & & & $D_{\rm s0}^*(2317) \bar D_{\rm s1}(2460)$ & 0.03 & $D_{\rm s}^* \bar D_{\rm s1}(2460)$ & 0.2 \\  
$D_{\rm s0}^*(2317) \bar D_{\rm s2}^*(2573)$ & 0.004$^\dag$ & & & $D_{\rm s0}^*(2317) \bar D_{\rm s1}(2536)$ & 0.1 & $D_{\rm s}^* \bar D_{\rm s1}(2536)$ & 0.8    \\  
$D_{\rm s} \bar D_{\rm s1}^*(2860)$ & 0.1$^\dag$ & & & $D_{\rm s0}^*(2317) \bar D_{\rm s2}^*(2573)$ & 0.1$^\dag$ & $D_{\rm s}^* \bar D_{\rm s2}^*(2573)$ & 4.0 \\ 
$D_{\rm s} \bar D_{\rm s3}^*(2860)$ & 2.1 &  &  & $D_{\rm s} \bar D_{\rm s1}^*(2860)$ & 0.003$^\dag$ & $D_{\rm s0}^*(2317) \bar D_{\rm s0}^*(2317)$ & $2\cdot10^{-4}$ \\
$D_{\rm s} \bar D_{\rm s2}(1D_2)$ & $8\cdot10^{-4}$ &  &  & $D_{\rm s} \bar D_{\rm s3}^*(2860)$ & 1.6 & $D_{\rm s0}^*(2317) \bar D_{\rm s1}(2460)$ & 0.007 \\
$D_{\rm s} \bar D_{\rm s2}(1D_2')$ & 0.001 & & & $D_{\rm s} \bar D_{\rm s2}(1D_2)$ & 0.3 & $D_{\rm s0}^*(2317) \bar D_{\rm s1}(2536)$ & 0.1 \\  
& & & & $D_{\rm s} \bar D_{\rm s2}(1D_2')$ & 0.2 & $D_{\rm s0}^*(2317) \bar D_{\rm s2}^*(2573)$ & 0.2$^\dag$ \\   
 &  &  &  & & & $D_{\rm s1}(2460) \bar D_{\rm s1}(2460)$ & 0.2 \\  
 &  &  &  & & & $D_{\rm s} \bar D_{\rm s1}^*(2860)$ & 0.04$^\dag$ \\
 &  &  &  & & & $D_{\rm s} \bar D_{\rm s3}^*(2860)$ & 0.8$^\dag$ \\
 &  &  &  & & & $D_{\rm s} \bar D_{\rm s2}(1D_2)$ & 0.2 \\
 &  &  &  & & & $D_{\rm s} \bar D_{\rm s2}(1D_2')$ & 0.1 \\
\hline
tot open-flavor & 187 & tot open-flavor & 157 & tot open-flavor & 201 & tot open-flavor & 183 \\                                                                                                                                                                                                                                                                                                                                                                                                                                                                                                                                                                                                            
\end{tabular}
\end{ruledtabular}
\caption{As Table \ref{tab:ChiC(4P)-strong}, but for $\chi_{\rm c}(5P)$ states. All the widths are in MeV.}
\label{tab:ChiC(5P)-strong}  
\end{table*}

\begin{table*}
\footnotesize
\centering
\begin{ruledtabular}
\begin{tabular}{cc|cc|cc} 
$\chi_{\rm c0}(4P)$ as $\chi_{\rm c0}(4500)$ & Width & $\chi_{\rm c0}(4P)$ as $\chi_{\rm c0}(4700)$ & Width & $\chi_{\rm c0}(5P)$ as $\chi_{\rm c0}(4700)$ & Width \\
decay channel &           [MeV]                            & decay channel &      [MeV]               &  decay channel &    [MeV]   \\                           
\hline                                     
$D \bar D$ & 1.0 & $D \bar D$ & 5.9 & $D \bar D$ & 3.2   \\ 
$D^* \bar D^*$ & 22.6 & $D^* \bar D^*$ & 1.7 & $D^* \bar D^*$ & 4.9    \\    
$D \bar D_0(2550)$ & 1.5$^\dag$ & $D \bar D_0(2550)$ & 9.8 & $D \bar D_0(2550)$ & 0.004$^\dag$   \\        
$D \bar D_1(2420)$ & 5.5 & $D^* \bar D_1(2^3S_1)$ & 54.8 & $D^* \bar D_1(2^3S_1)$ & 33.6   \\ 
$D \bar D_1(2430)$ & 2.2 & $D \bar D_1(2420)$ & 15.3 & $D \bar D_1(2420)$ & 0.01$^\dag$     \\   
$D^* \bar D_0^*(2300)$ & 0.2$^\dag$ & $D \bar D_1(2430)$ & 4.8 & $D \bar D_1(2430)$ & 0.04$^\dag$    \\                                                                                    
$D^* \bar D_1(2420)$ & 9.9 & $D^* \bar D_0^*(2300)$ & 3.9 & $D^* \bar D_0^*(2300)$ & 0.1$^\dag$ \\                                                                                                                                                                  
$D^* \bar D_1(2430)$ & 28.4 & $D^* \bar D_1(2420)$ & 6.4 & $D^* \bar D_1(2420)$ & 2.3    \\                                                                                                                                          
$D^* \bar D_2^*(2460)$ & 6.1 & $D^* \bar D_1(2430)$ & 14.4 & $D^* \bar D_1(2430)$ & 7.6 \\                                                                                                                                                                                                                                                                                                                                                                           
$D_{\rm s} \bar D_{\rm s}$ & 1.2 & $D^* \bar D_2^*(2460)$ & 87.2 & $D^* \bar D_2^*(2460)$ & 18.3   \\                                                                                                                                                                                                                                                                                                                                                                         
$D_{\rm s}^* \bar D_{\rm s}^*$ & 0.2 & $D_0^*(2300) \bar D_0^*(2300)$ & 0.004$^\dag$ & $D_0^*(2300) \bar D_0^*(2300)$ & 0.2$^\dag$ \\       
$D_{\rm s} \bar D_{\rm s1}(2460)$ & 8.6 & $D \bar D_2(1D_2)$ & 1.4 & $D \bar D_2(1D_2)$ & 1.4   \\ 
$D_{\rm s} \bar D_{\rm s1}(2536)$ & 0.03 & $D \bar D_2(1D_2')$ & 0.9 & $D \bar D_2(1D_2')$ & 0.9   \\ 
$D_{\rm s}^* \bar D_{\rm s0}^*(2317)$ & 1.7 & $D_{\rm s} \bar D_{\rm s}$ & 0.4 & $D_{\rm s} \bar D_{\rm s}$ & 0.1 \\ 
& & $D_{\rm s}^* \bar D_{\rm s}^*$ & 4.0 & $D_{\rm s}^* \bar D_{\rm s}^*$ & 0.8   \\ 
& & $D_{\rm s} \bar D_{\rm s0}(2^1S_0)$ & 2.5 & $D_{\rm s} \bar D_{\rm s0}(2^1S_0)$ & 0.3    \\                    
& & $D_{\rm s} \bar D_{\rm s1}(2460)$ & 3.1 & $D_{\rm s} \bar D_{\rm s1}(2460)$ & 2.2    \\ 
& & $D_{\rm s} \bar D_{\rm s1}(2536)$ & 0.05 & $D_{\rm s} \bar D_{\rm s1}(2536)$ & 0.09  \\ 
& & $D_{\rm s}^* \bar D_{\rm s0}^*(2317)$ & 0.5 & $D_{\rm s}^* \bar D_{\rm s0}^*(2317)$ & 0.5  \\ 
& & $D_{\rm s}^* \bar D_{\rm s1}(2460)$ & 0.2 & $D_{\rm s}^* \bar D_{\rm s1}(2460)$ & 0.003 \\ 
& & $D_{\rm s}^* \bar D_{\rm s1}(2536)$ & 7.6 & $D_{\rm s}^* \bar D_{\rm s1}(2536)$ & 3.2 \\ 
& & $D_{\rm s}^* \bar D_{\rm s2}^*(2573)$ & 0.6$^\dag$ & $D_{\rm s}^* \bar D_{\rm s2}^*(2573)$ & 0.3$^\dag$ \\ 
& & $D_{\rm s0}^*(2317) \bar D_{\rm s0}^*(2317)$ & 0.02 & $D_{\rm s0}^*(2317) \bar D_{\rm s0}^*(2317)$ & 0.005 \\   
\hline
tot open-flavor & 89 & tot open-flavor & 225 & tot open-flavor & 80 \\                                                                                                                                                                                                                                                                                                                                                                                                                                                                                                                                                                                                            
\end{tabular}
\end{ruledtabular}
\caption{As Table \ref{tab:ChiC(4P)-strong}, but for the decays of $\chi_{\rm c0}(4P)$ as $\chi_{\rm c0}(4500)$ or $\chi_{\rm c0}(4700)$ and $\chi_{\rm c0}(5P)$ as $\chi_{\rm c0}(4700)$. Here, we use the experimental values of the $\chi_{\rm c0}(4500)$ and $\chi_{\rm c0}(4700)$ meson masses \cite{Zyla:2020zbs}.}
\label{tab:X(4500)-X(4700)-strong}  
\end{table*}

Our theoretical results, obtained by using the pair-creation model parameters of Table \ref{tab:3P0parameters}, are given in Tables \ref{tab:ChiC(4P)-strong}-\ref{tab:X(4500)-X(4700)-strong}.
It is worth to note that: I) the calculated total open-charm strong decay widths of $\chi_{\rm c}(4P)$s and $\chi_{\rm c}(5P)$s of Tables \ref{tab:ChiC(4P)-strong} and \ref{tab:ChiC(5P)-strong} are quite large; they are in the order of $150-200$ MeV. 
If we make the hypothesis of considering the open-charm as the largely dominant decay modes of higher charmonia, a comparison with the existing and forthcoming experimental data can be easily done.
If our pair-creation model results are confirmed by the future experiment data, the $\chi_{\rm c}(4P,5P)$ states will be reasonably interpreted as charmonium (or charmonium-like) states dominated by the $c \bar c$ component; II) the results of Table \ref{tab:X(4500)-X(4700)-strong}, obtained by making the tentative assignments $\chi_{\rm c0}(4500) \rightarrow \chi_{\rm c0}(4P)$ and $\chi_{\rm c0}(4700) \rightarrow \chi_{\rm c0}(4P)$ or $\chi_{\rm c0}(5P)$, seem to span a wider interval. In particular, one can notice that the assignments $\chi_{\rm c0}(4700) \rightarrow \chi_{\rm c0}(4P)$ and $\chi_{\rm c0}(4700) \rightarrow \chi_{\rm c0}(5P)$ produce results for the total open-flavor widths of 225 and 80 MeV, respectively.
A comparison with the total experimental width of the $\chi_{\rm c0}(4700)$ \cite{Zyla:2020zbs}, $120\pm31^{+42}_{-33}$ MeV, seems to favor the $\chi_{\rm c0}(5P)$ assignment, even though the experimental error is so large that it is difficult to draw a definitive conclusion. Our result for the total open-flavor width of the $\chi_{\rm c0}(4500)$ as $\chi_{\rm c0}(4P)$, 89 MeV, is in good accordance with the experimental total decay width of the $\chi_{\rm c0}(4500)$, $92\pm21^{+21}_{-20}$ MeV.
In light of this, our $^3P_0$ model results would suggest the assignments $\chi_{\rm c0}(4500) \rightarrow \chi_{\rm c0}(4P)$ and $\chi_{\rm c0}(4700) \rightarrow \chi_{\rm c0}(5P)$, even though $\chi_{\rm c0}(4700) \rightarrow \chi_{\rm c0}(4P)$ cannot be ruled out completely; III) there are decay channels whose widths change notably by switching from a specific assignment to another; see e.g. the $D^* \bar D^*$ and $D^* \bar D_2^*(2460)$ decay mode results from Table \ref{tab:X(4500)-X(4700)-strong}.
Therefore, a detailed study of the $D^* \bar D^*$, $D^* \bar D_2^*(2460)$, $D^* \bar D_1(2^3S_1)$ ... decay channels may help considerably in the assignment procedure.

Finally, it is interesting to discuss, in the context of a $^3P_0$ model calculation, the possible importance of: I) averaging the open-flavor widths of charmonia over the Breit-Wigner distributions of the daughter mesons. One can observe that, in the present study, the decay widths into charmed meson pairs do not take the widths of the final states into account. However, these are sizable, $\mathcal O(100 \mbox{ MeV})$, for several of the decays discussed here, and may thus affect some of the results; see e.g. the $D_0^*(2300)$, whose width is $274 \pm 40$ MeV, and the $D_0(2550)$, whose width is $135 \pm 17$ MeV \cite{Zyla:2020zbs}. 
There are even cases of charmed-strange mesons whose width is large, like the $D_{s1}^*(2860)$. However, the contribution of the charmed-strange meson decay channels to the total widths of charmonia is expected to be smaller because of the effective pair-creation strength suppression mechanism of Eq. (\ref{eqn:gamma0-eff}).
In light of this, we conclude that some of our results for the open-flavor strong decay widths of $\chi_{\rm c}(4P,5P)$ states may not be reliable. In particular, this might the case of channels like $h_{\rm c}(4P) \rightarrow D \bar D_1 (2420)$ or $D \bar D_1 (2430)$, whose calculated widths are small but they could be larger once the effects of averaging over the widths of the final states are taken into account. 
In conclusion, we believe that it would be interesting to see how our results for the open-flavor strong decay widths of charmonia will change after this averaging procedure is performed.
This will be the subject of a subsequent paper \cite{TBP}; II) including the quark form factor (QFF) in the $^3P_0$ model transition operator; see Appendix \ref{3P0model}. The QFF was not considered in the original formulation of the $^3P_0$ model \cite{Micu,LeYaouanc}, but it was introduced in a second stage with the phenomenological purpose to take the effective size of the $q \bar q$ pair of created quarks into account \cite{bottomonium,Geiger:1989yc,Bijker:2009up,Strong2015}.
Its possible importance in our results can be somehow quantified by calculating the widths of some specific decay channels, like  $\psi(3770) \rightarrow D \bar D$, by means of the standard $^3P_0$ model transition operator and the modified one, which includes the quark form factor. In the former case, we get $\Gamma[\psi(3770) \rightarrow D \bar D] = 27$ MeV; in the latter, we obtain $\Gamma[\psi(3770) \rightarrow D \bar D] = 80$ MeV. 
The second result for the width, i.e. 80 MeV, is outsize. It is clear that realistic results for the open-flavor strong decay widths of charmonia can be obtained in both cases; however, if the QFF is not taken into account, the values of the model parameters of Table \ref{tab:3P0parameters} need to be re-fitted to the data; III) extracting a different value of the harmonic oscillator (h.o.) parameter for each state involved in the decays rather than using a single value for them all, as it is done here. The former approach was used e.g. in Refs. \cite{Kokoski:1985is,Ko:2000jx,Kumar:2011ff}.
Consider, in particular, the prescriptions of Ref. \cite{Ko:2000jx}. There, the h.o. parameters of charmonia were fitted to their squared radii from potential model calculations \cite{Eichten:1978tg}. In the case of the $J/\psi$, $\psi(2S)$ and $\psi(3770)$, the authors got $\alpha_{\rm ho} = 0.52, 0.39$ and 0.37 GeV, respectively.
Furthermore, the value of $\alpha_{\rm ho}$ for $D$ mesons (and that of $\gamma_0$) were fitted to the open-charm decays of the $\psi(3770)$ and $\psi(4040)$.
The main advantage of the previous approach with respect to that used in the present paper resides in the possibility of obtaining results for the decays based on more realistic wave functions for the parent mesons. On the contrary, the prescriptions used here have the advantage of a greater flexibility and of a smaller number of free parameters; IV) finally, we have to comment that a realistic value of $\gamma_0$ can be found in the range $0.3-0.5$, approximately. See e.g. Ref. \cite{Ko:2000jx}, where $\gamma_0 = 0.35$ was fitted to the open-charm decays of the $\psi(3770)$ and $\psi(4040)$, and Ref. \cite{Barnes:2005pb}, where a value of $\gamma_0 = 0.4$ made it possible to obtain a good reproduction of the open-charm decay widths of charmonia up to $2F$ and $1G$ resonances.
The value used here and in Refs. \cite{Ferretti:2013faa,Ferretti:2015rsa,Ferretti:2014xqa}, $\gamma_0 = 0.510$ (see Table \ref{tab:3P0parameters}), was fitted to the strong decay widths of $3S$, $2P$, $1D$, and $2D$ charmonia.
This value is different from those used in other studies \cite{Ko:2000jx,Barnes:2005pb} because of the presence here of the QFF and of different choices of $\alpha_{\rm ho}$.
Evidently, all the model parameter values are tightly connected to one another: changing the value of one of them will automatically require a redefinition of the values of all the other model parameters or, at least, of a part of them.

\section{Threshold mass-shifts of $\chi_{\rm c}(4P,5P)$ states in a coupled-channel model}
\label{Threshold mass-shifts}
Here, we make use of the UQM-based CCM of Refs. \cite{Ferretti:2018tco,Ferretti:2020civ} to explore the possible assignments $\chi_{\rm c0}(4P) \rightarrow \chi_{\rm c0}(4500)$ or $\chi_{\rm c0}(4700)$ and $\chi_{\rm c0}(5P) \rightarrow \chi_{\rm c0}(4700)$. To do that, we calculate the threshold corrections to the bare masses of the $\chi_{\rm c0}(4500)$ and $\chi_{\rm c0}(4700)$ resonances to see if the introduction of loop effects can help to reconcile the relativized QM \cite{Godfrey:1985xj} results for $\chi_{\rm c0}(4P,5P)$ states, see Table \ref{tab:bare-masses} herein, with the experimental data \cite{Zyla:2020zbs,Aaij:2016iza,Aaij:2016nsc}.

In the UQM, \cite{Heikkila:1983wd,Geiger:1989yc,bottomonium,Bijker:2009up,Pennington:2007xr,Lu:2016mbb,Ortega:2012rs} the wave function of a hadron,
\begin{equation}	
	\footnotesize
	\label{eqn:Psi-A}
	\begin{array}{l}	
	\left| \psi_A \right\rangle = {\cal N} \left[ \left| A \right\rangle + \displaystyle \sum_{BC} \int k^2 dk \, 
	\left| BC k \, \ell J \right\rangle \frac{ \left\langle BC k \, \ell J \right| T^{\dagger} \left| A \right\rangle}{M_A - E_B - E_C} \right] \mbox{ }, 
	\end{array}
\end{equation}
is the superposition of a valence core, $\left| A \right\rangle = \left| Q \bar Q \right\rangle$, plus higher Fock components, $\left| BC \right\rangle = \left| Q \bar q; q \bar Q \right\rangle$, due to the creation of light $q \bar q$ pairs.
The sum is extended over a complete set of meson-meson intermediate states $\left| BC \right\rangle$ and the amplitudes, $\left\langle BC q \, \ell J \right| T^{\dagger} \left| A \right\rangle$, are computed within the $^3P_0$ pair-creation model of Sec. \ref{3P0-formulas}.

The physical masses of hadrons are calculated as 
\begin{equation}
	\label{eqn:Ma-UQM}
	M_A^{\rm UQM} = E_A + \Sigma(M_A) \mbox{ }.
\end{equation}
Here, $E_A$ is the bare mass of the hadron $A$, and
\begin{equation}
	\label{eqn:self-a}
	\Sigma(M_A) = \sum_{BC} \int_0^{\infty} k^2 dk \mbox{ } 
	\frac{\left| \left\langle BC k  \, \ell J \right| T^\dag \left| A \right\rangle \right|^2}{M_A - E_B(k) - E_C(k)}  
\end{equation}
is a self-energy correction. The bare masses $E_A$ are usually computed in a potential model, whose parameters  are fixed by fitting Eq. (\ref{eqn:Ma-UQM}) to the reproduction of the experimental data; see e.g. Refs. \cite{Ferretti:2013faa,Ferretti:2013vua}.
\begin{table}[htbp]
\centering
\begin{ruledtabular}
\begin{tabular}{ccccc} 
State                        & $E_A$             & $\Sigma(M_A) - \Delta_{\rm th}$           & $M_A^{\rm th}$           & $M_A^{\rm exp}$ \\
                                &             [MeV]  &                                       [MeV] &  [MeV]                        &  [MeV] \\
\hline
$h_{\rm c}(4P)$       & 4634               & $-33^\dag$; $-21^\$$ & $4601^\dag$; $4613^\$$ & -- \\
$X(4500)$               & 4613               & $0^\dag$; $0^\$$  & $4613^\dag$; $4613^\$$ & $4506\pm11^{+12}_{-15}$ \\
$\chi_{\rm c1}(4P)$ & 4633               & $-23^\dag$; $-19^\$$  & $4610^\dag$; $4614^\$$ & --  \\
$\chi_{\rm c2}(4P)$ & 4650               & $-50^\dag$; $-33^\$$  & $4600^\dag$; $4617^\$$ & --  \\  
\hline
$h_{\rm c}(4P)$       & 4634               & $-10^\dag$; $-12^\$$ & $4624^\dag$; $4622^\$$ & -- \\
$X(4700)$               & 4613               & $-17^\dag$; $0^\$$  & $4596^\dag$; $4613^\$$ & $4704\pm10^{+14}_{-24}$ \\
$\chi_{\rm c1}(4P)$ & 4633               & $0^\dag$; $-10^\$$  & $4633^\dag$; $4623^\$$ & --  \\
$\chi_{\rm c2}(4P)$ & 4650               & $-10^\dag$; $-24^\$$  & $4640^\dag$; $4626^\$$ & --  \\     
\hline
$h_{\rm c}(5P)$       & 4919 & $-23^\dag$; $-30^\$$ & $4896^\dag$; $4889^\$$ & -- \\
$X(4700)$               & 4902 & $0^\dag$; $0^\$$  & $4902^\dag$; $4902^\$$ & $4704\pm10^{+14}_{-24}$ \\
$\chi_{\rm c1}(5P)$ & 4919 & $-22^\dag$; $-30^\$$  & $4897^\dag$; $4889^\$$ & --  \\
$\chi_{\rm c2}(5P)$ & 4934 & $-24^\dag$; $-33^\$$  & $4910^\dag$; $4901^\$$ & --  \\                                                      
\end{tabular}
\end{ruledtabular}
\caption{Coupled-channel model results for the relative threshold corrections of $\chi_{\rm c}(4P)$ and $\chi_{\rm c}(5P)$ states, calculated via Eq. (\ref{eqn:new-Ma}). The self-energies $\Sigma(M_A)$ are extracted from Tables \ref{tab:Mass-shifts-4P} and \ref{tab:Mass-shifts-5P}. In the $\chi_{\rm c0}(4P)$ case, we try the assignments $\chi_{\rm c0}(4P) \rightarrow \chi_{\rm c0}(4500)$ (top part of the table) and $\chi_{\rm c0}(4P) \rightarrow \chi_{\rm c0}(4700)$ (in the middle); in the $\chi_{\rm c0}(5P)$ case, we only consider the assignment $\chi_{\rm c0}(5P) \rightarrow \chi_{\rm c0}(4700)$ (bottom part of the table). The results marked by the superscript $\dag$ are obtained by considering $1S2S$ and $1P1P$ loop contributions, those marked by $\$$ by including $1S2S$, $1P1P$ and also $1S1P$ loop contributions.}
\label{tab:ChiC(3P)-splittings}  
\end{table}

The idea at the basis of the coupled-channel approach of Refs.~\cite{Ferretti:2018tco,Ferretti:2020civ} is slightly different.
There, one can study a single multiplet at a time, like $\chi_{\rm c}(2P)$ or $\chi_{\rm b}(3P)$, without the need of considering an entire meson sector to re-fit the potential model parameters to the reproduction of the physical masses of Eq. (\ref{eqn:Ma-UQM}).
This is because the bare masses $E_A$ are directly extracted from the relativized QM predictions of Refs.~\cite{Barnes:2005pb,Godfrey:1985xj}; see Table \ref{tab:bare-masses}. 
In our coupled-channel model approach, the physical masses of the meson multiplet members are given by \cite{Ferretti:2018tco,Ferretti:2020civ}
\begin{equation}
	\label{eqn:new-Ma}
	M_A^{\rm CCM} = E_A + \Sigma(M_A) + \Delta_{\rm th} \mbox{ },
\end{equation}
where $E_A$ and $\Sigma(M_A)$ have the same meaning as in Eq. (\ref{eqn:Ma-UQM}) and $\Delta_{\rm th}$ is a parameter.
\begin{table*}
\footnotesize
\centering
\begin{ruledtabular}
\begin{tabular}{c|c|c|c|c|c} 
State                & $D_0^*(2300) \bar D_0^*(2300)$ & $D_0^*(2300) \bar D_1(2420)$  & $D_0^*(2300) \bar D_1(2430)$ & $D_0^*(2300) \bar D_2^*(2460)$ & $D_1(2420) \bar D_1(2420)$  \\
\hline
$h_c(4P)$          & -- & $-8.2$ & $-3.8$ & $-0.05$ & -- \\
$\chi_{\rm c0}(4P)$ as $X(4500)$ & $-4.6$ & -- & -- & $-4.1$ & $-6.7$ \\
$\chi_{\rm c0}(4P)$ as $X(4700)$ & $-5.2$ & -- & -- & $-4.8$ & $-8.8$ \\
$\chi_{\rm c1}(4P)$ & -- & $-2.0$ & $-7.3$ & $-3.6$ & $-3.7$ \\
$\chi_{\rm c2}(4P)$ & $-1.2$ & $-1.1$ & $-2.9$ & $-4.9$ & $-6.1$ \\
\hline
State                & $D_1(2420) \bar D_1(2430)$ & $D_1(2420) \bar D_2^*(2460)$  & $D_1(2430) \bar D_1(2430)$ & $D_1(2430) \bar D_2^*(2460)$ & $D_2^*(2460) \bar D_2^*(2460)$ \\
\hline
$h_c(4P)$          & $-5.5$ & $-18.8$ & $-5.6$ & $-12.1$ & $-9.8$ \\
$\chi_{\rm c0}(4P)$ as $X(4500)$ & $-9.7$ & $-3.1$ & $-7.7$ & $-7.6$ & $-13.5$ \\
$\chi_{\rm c0}(4P)$ as $X(4700)$ & $-12.3$ & $-3.9$ & $-9.1$ & $-8.9$ & $-16.7$ \\
$\chi_{\rm c1}(4P)$ & $-8.3$ & $-13.3$ & $-3.1$ & $-13.2$ & $-9.5$ \\
$\chi_{\rm c2}(4P)$ & $-9.5$ & $-11.1$ & $-4.7$ & $-9.6$ & $-13.6$ \\
\hline
State                & $D_{\rm s0}^*(2317) \bar D_{\rm s0}^*(2317)$ & $D_{\rm s0}^*(2317) \bar D_{\rm s1}(2460)$  & $D_{\rm s0}^*(2317) \bar D_{\rm s1}(2536)$ & $D_{\rm s0}^*(2317) \bar D_{\rm s2}^*(2573)$ & $D_{\rm s1}(2460) \bar D_{\rm s1}(2460)$ \\
\hline
$h_c(4P)$          & -- & $-0.7$ & $-1.0$ & $-0.02$ & -- \\
$\chi_{\rm c0}(4P)$ as $X(4500)$ & $-0.7$ & -- & -- & $-0.5$ & $-0.9$ \\
$\chi_{\rm c0}(4P)$ as $X(4700)$ & $-0.8$ & -- & -- & $-0.6$ & $-1.2$ \\
$\chi_{\rm c1}(4P)$ & -- & $-0.5$ & $-0.9$ & $-0.4$ & $-0.7$ \\
$\chi_{\rm c2}(4P)$ & $-0.2$ & $-0.3$ & $-0.4$ & $-0.4$ & $-1.0$ \\
\hline
State                & $D_{\rm s1}(2460) \bar D_{\rm s1}(2536)$ & $D_{\rm s1}(2460) \bar D_{\rm s2}^*(2573)$  & $D_{\rm s1}(2536) \bar D_{\rm s1}(2536)$ & $D_{\rm s1}(2536) \bar D_{\rm s2}^*(2573)$ & $D_{\rm s2}^*(2573) \bar D_{\rm s2}^*(2573)$ \\
\hline
$h_c(4P)$          & $-1.5$ & $-2.8$ & $-0.5$ & $-1.4$ & $-1.8$ \\
$\chi_{\rm c0}(4P)$ as $X(4500)$ & $-1.8$ & $-0.6$ & $-1.0$ & $-1.2$ & $-2.2$ \\
$\chi_{\rm c0}(4P)$ as $X(4700)$ & $-2.1$ & $-0.7$ & $-1.1$ & $-1.3$ & $-2.6$ \\
$\chi_{\rm c1}(4P)$ & $-0.8$ & $-2.1$ & $-0.8$ & $-1.8$ & $-1.6$ \\
$\chi_{\rm c2}(4P)$ & $-1.2$ & $-1.9$ & $-0.8$ & $-1.2$ & $-2.3$ \\
\hline
State                & $D \bar D_0(2550)$ & $D \bar D_1(2^3S_1)$ & $D^* \bar D_0(2550)$ & $D^* \bar D_1(2^3S_1)$ & $D_{\rm s} \bar D_{\rm s0}(2^1S_0)$ \\
\hline
$h_c(4P)$          & -- & $-8.4$ & $-12.6$ & $-29.6$ & -- \\
$\chi_{\rm c0}(4P)$ as $X(4500)$ & $-0.3$ & -- & -- & $-26.6$ & $-1.7$ \\
$\chi_{\rm c0}(4P)$ as $X(4700)$ & 2.0 & -- & -- & $-55.4$ & 1.0 \\
$\chi_{\rm c1}(4P)$ & -- & $-9.4$ & $-4.4$ & $-26.5$ & -- \\
$\chi_{\rm c2}(4P)$ & 1.1 & $-2.3$ & $-10.8$ & $-54.1$ & $-1.3$ \\
\hline
State                & $D_{\rm s} \bar D_{\rm s1}(2^3S_1)$ & $D_{\rm s}^* \bar D_{\rm s0}(2^1S_0)$ & $D_{\rm s}^* \bar D_{\rm s1}(2^3S_1)$ & $D \bar D_0^*(2300)$ & $D \bar D_1(2420)$  \\
\hline
$h_c(4P)$          & $-2.6$ & $-1.8$ & $-3.5$ & $-8.0$ & -- \\
$\chi_{\rm c0}(4P)$ as $X(4500)$ & -- & -- & $-4.5$ & -- & $-10.7$ \\
$\chi_{\rm c0}(4P)$ as $X(4700)$ & -- & -- & $-6.2$ & -- & 3.5 \\
$\chi_{\rm c1}(4P)$ & $-2.6$ & $-1.7$ & $-3.6$ & $-0.2$ & $-7.5$ \\
$\chi_{\rm c2}(4P)$ & $-1.6$ & $-1.2$ & $-4.1$ & -- & $-2.3$ \\
\hline
State                & $D \bar D_1(2430)$ & $D \bar D_2^*(2460)$ & $D^* \bar D_0^*(2300)$ & $D^* \bar D_1(2420)$ & $D^* \bar D_1(2430)$ \\
\hline
$h_c(4P)$          & $-0.2$ & $-4.8$ & $-0.2$ & $-20.9$ & $-23.0$ \\
$\chi_{\rm c0}(4P)$ as $X(4500)$ & $-15.3$ & -- & $-5.7$ & $-12.9$ & $-5.7$ \\
$\chi_{\rm c0}(4P)$ as $X(4700)$ & $-15.4$ & -- & $-4.9$ & $-15.9$ & $-12.2$ \\
$\chi_{\rm c1}(4P)$ & $-4.7$ & $-4.7$ & $-7.2$ & $-15.6$ & $-18.7$ \\
$\chi_{\rm c2}(4P)$ & $-5.6$ & $-1.4$ & $-10.2$ & $-11.4$ & $-14.5$ \\
\hline
State                & $D^* \bar D_2^*(2460)$ & $D_{\rm s} \bar D_{\rm s0}^*(2317)$ & $D_{\rm s} \bar D_{\rm s1}(2460)$ & $D_{\rm s} \bar D_{\rm s1}(2536)$ & $D_{\rm s} \bar D_{\rm s2}^*(2573)$ \\
\hline
$h_c(4P)$          & $-20.2$ & $-0.5$ & -- & $-0.06$ & $-2.7$ \\
$\chi_{\rm c0}(4P)$ as $X(4500)$ & $-43.5$ & -- & $-0.4$ & $-2.1$ & -- \\
$\chi_{\rm c0}(4P)$ as $X(4700)$ & $-13.2$ & -- & $-2.0$ & $-2.0$ & -- \\
$\chi_{\rm c1}(4P)$ & $-29.3$ & $-0.05$ & $-1.0$ & 0.4 & $-1.3$ \\
$\chi_{\rm c2}(4P)$ & $-25.7$ & -- & $-1.0$ & $-0.9$ & $-1.7$ \\
\hline
State                & $D_{\rm s}^* \bar D_{\rm s0}^*(2317)$ & $D_{\rm s}^* \bar D_{\rm s1}(2460)$ & $D_{\rm s}^* \bar D_{\rm s1}(2536)$ & $D_{\rm s}^* \bar D_{\rm s2}^*(2573)$ & Total \\
\hline
$h_c(4P)$          & $-0.05$ & $-4.0$ & $-3.3$ & $-4.7$ & $-132.1^\dag$; $-224.8^\$$ \\
$\chi_{\rm c0}(4P)$ as $X(4500)$ & $-0.2$ & $-1.8$ & $-1.2$ & $-5.3$ & $-99.0^\dag$; $-203.8^\$$ \\
$\chi_{\rm c0}(4P)$ as $X(4700)$ & $-0.4$ & $-1.6$ & $-1.4$ & $-8.5$ & $-138.7^\dag$; $-212.7^\$$ \\
$\chi_{\rm c1}(4P)$ & $-0.3$ & $-2.8$ & $-2.8$ & $-5.0$ & $-121.8^\dag$; $-222.6^\$$ \\
$\chi_{\rm c2}(4P)$ & $-0.6$ & $-2.7$ & $-2.2$ & $-7.4$ & $-148.7^\dag$; $-236.3^\$$ \\
\end{tabular}
\end{ruledtabular}
\caption{Self-energy corrections, $\Sigma(M_A)$ (in MeV), to the bare masses of $\chi_{\rm c}(4P)$ states, calculated via Eq. (\ref{eqn:self-a}). The values of the UQM parameters are extracted from \cite[Table II]{Ferretti:2013faa}. The contributions of those channels denoted by -- are suppressed by selection rules. In the case of the $\chi_{\rm c0}(4P)$, we provide results for both the $\chi_{\rm c0}(4P) \rightarrow X(4500)$ and $\chi_{\rm c0}(4P) \rightarrow X(4700)$ assignments. The total self-energies marked by the superscript $\dag$ are the sum of $1S2S$ and $1P1P$ loop contributions, those marked by $\$$ are the sum of $1S2S$, $1P1P$ and also $1S1P$ loop contributions.} 
\label{tab:Mass-shifts-4P}  
\end{table*}
For each multiplet we consider, this is the only free parameter of our calculation.
It is defined as the smallest self-energy correction (in terms of absolute value) among those of the multiplet members; see \cite[Secs. 2.2 and 2.3]{Ferretti:2018tco} and \cite[Secs. IIB and IIIC]{Ferretti:2020civ}.
The introduction of $\Delta_{\rm th}$ in Eq. (\ref{eqn:new-Ma}) represents our ``renormalization'' or ``subtraction'' prescription for the threshold mass-shifts in the UQM.
The UQM model parameters, which we need in the calculation of the $\left\langle BC k  \, \ell J \right| T^\dag \left| A \right\rangle$ vertices and the self-energies of Eq. (\ref{eqn:self-a}), are reported in Table \ref{tab:3P0parameters}. See also Appendix \ref{3P0model}.

\begin{table*}
\footnotesize
\centering
\begin{ruledtabular}
\begin{tabular}{c|c|c|c|c|c} 
State                & $D_0^*(2300) \bar D_0^*(2300)$ & $D_0^*(2300) \bar D_1(2420)$  & $D_0^*(2300) \bar D_1(2430)$ & $D_0^*(2300) \bar D_2^*(2460)$ & $D_1(2420) \bar D_1(2420)$ \\
\hline
$h_c(5P)$          & -- & $-5.0$ & $-3.7$ & $-0.02$ & -- \\
$\chi_{\rm c0}(5P)$ as $X(4700)$ & $-3.6$ & -- & -- & $-3.4$ & $-6.3$ \\
$\chi_{\rm c1}(5P)$ & -- & $-1.9$ & $-5.3$ & $-2.0$ & $-3.0$ \\
$\chi_{\rm c2}(5P)$ & $-1.1$ & $-1.1$ & $-2.3$ & $-3.2$ & $-6.5$ \\
\hline
State                & $D_1(2420) \bar D_1(2430)$ & $D_1(2420) \bar D_2^*(2460)$  & $D_1(2430) \bar D_1(2430)$ & $D_1(2430) \bar D_2^*(2460)$ & $D_2^*(2460) \bar D_2^*(2460)$ \\
\hline
$h_c(5P)$          & $-5.4$ & $-25.1$ & $-4.5$ & $-14.4$ & $-11.3$ \\
$\chi_{\rm c0}(5P)$ as $X(4700)$ & $-7.4$ & $-2.4$ & $-6.3$ & $-6.7$ & $-11.6$ \\
$\chi_{\rm c1}(5P)$ & $-8.0$ & $-19.1$ & $-3.1$ & $-14.7$ & $-10.7$ \\
$\chi_{\rm c2}(5P)$ & $-9.1$ & $-13.3$ & $-4.6$ & $-9.3$ & $-19.8$ \\
\hline
State                & $D_{\rm s0}^*(2317) \bar D_{\rm s0}^*(2317)$ & $D_{\rm s0}^*(2317) \bar D_{\rm s1}(2460)$  & $D_{\rm s0}^*(2317) \bar D_{\rm s1}(2536)$ & $D_{\rm s0}^*(2317) \bar D_{\rm s2}^*(2573)$ & $D_{\rm s1}(2460) \bar D_{\rm s1}(2460)$ \\
\hline
$h_c(5P)$          & -- & $-0.4$ & $-0.8$ & $-0.02$ & -- \\
$\chi_{\rm c0}(5P)$ as $X(4700)$ & $-0.4$ & -- & -- & $-0.4$ & $-0.9$ \\
$\chi_{\rm c1}(5P)$ & -- & $-0.4$ & $-0.6$ & $-0.5$ & $-0.6$ \\
$\chi_{\rm c2}(5P)$ & $-0.2$ & $-0.3$ & $-0.3$ & $-0.2$ & $-1.0$ \\
\hline
State                & $D_{\rm s1}(2460) \bar D_{\rm s1}(2536)$ & $D_{\rm s1}(2460) \bar D_{\rm s2}^*(2573)$  & $D_{\rm s1}(2536) \bar D_{\rm s1}(2536)$ & $D_{\rm s1}(2536) \bar D_{\rm s2}^*(2573)$ & $D_{\rm s2}^*(2573) \bar D_{\rm s2}^*(2573)$ \\
\hline
$h_c(5P)$          & $-1.2$ & $-2.7$ & $-0.3$ & $-1.2$ & $-1.6$ \\
$\chi_{\rm c0}(5P)$ as $X(4700)$ & $-1.2$ & $-0.4$ & $-0.7$ & $-1.1$ & $-1.8$ \\
$\chi_{\rm c1}(5P)$ & $-0.8$ & $-2.1$ & $-0.6$ & $-1.5$ & $-1.5$ \\
$\chi_{\rm c2}(5P)$ & $-1.0$ & $-1.7$ & $-0.7$ & $-1.1$ & $-2.1$ \\
\hline
State                & $D_0(2550) \bar D_0^*(2300)$ & $D_0(2550) \bar D_1(2420)$ & $D_0(2550) \bar D_1(2430)$ & $D_0(2550) \bar D_2^*(2460)$ & $D_1(2^3S_1) \bar D_0^*(2300)$ \\
\hline
$h_c(5P)$          & $-1.2$ & -- & $-0.03$ & $-4.3$ & $-0.03$ \\
$\chi_{\rm c0}(5P)$ as $X(4700)$ & -- & $-2.6$ & $-1.4$ & -- & $-1.0$ \\
$\chi_{\rm c1}(5P)$ & $-0.01$ & $-1.1$ & $-2.1$ & $-3.2$ & $-1.7$ \\
$\chi_{\rm c2}(5P)$ & -- & $-2.5$ & $-1.8$ & $-2.1$ & $-1.9$ \\
\hline
State                & $D_1(2^3S_1) \bar D_1(2420)$ & $D_1(2^3S_1) \bar D_1(2430)$ & $D_1(2^3S_1) \bar D_2^*(2460)$ & $D_{\rm s}(2^1S_0) \bar D_{\rm s0}^*(2317)$ & $D_{\rm s}(2^1S_0) \bar D_{\rm s1}(2460)$ \\
\hline
$h_c(5P)$          & $-5.2$ & $-4.6$ & $-5.1$ & $-0.3$ & -- \\
$\chi_{\rm c0}(5P)$ as $X(4700)$ & $-1.3$ & $-2.2$ & $-5.5$ & -- & $-0.4$ \\
$\chi_{\rm c1}(5P)$ & $-4.4$ & $-3.2$ & $-5.8$ & $-0.01$ & $-0.2$ \\
$\chi_{\rm c2}(5P)$ & $-3.3$ & $-3.1$ & $-7.3$ & -- & $-0.4$ \\
\hline
State                 & $D_{\rm s}(2^1S_0) \bar D_{\rm s1}(2536)$ & $D_{\rm s}(2^1S_0) \bar D_{\rm s2}^*(2573)$ & $D_{\rm s1}^*(2700) \bar D_{\rm s0}^*(2317)$ & $D_{\rm s1}^*(2700) \bar D_{\rm s1}(2460)$ & $D_{\rm s1}^*(2700) \bar D_{\rm s1}(2536)$ \\
\hline
$h_c(5P)$          & $-0.01$ & $-0.6$ & $-0.01$ & $-0.9$ & $-0.7$ \\
$\chi_{\rm c0}(5P)$ as $X(4700)$ & $-0.3$ & -- & $-0.2$ & $-0.3$ & $-0.3$ \\
$\chi_{\rm c1}(5P)$ & $-0.2$ & $-0.4$ & $-0.2$ & $-0.7$ & $-0.5$  \\
$\chi_{\rm c2}(5P)$ & $-0.2$ & $-0.3$ & $-0.3$ & $-0.6$ & $-0.5$ \\
\hline
State                & $D_{\rm s1}^*(2700) \bar D_{\rm s2}^*(2573)$ &  &  &  & Total \\
\hline
$h_c(5P)$          & $-0.9$ & & & & $-77.6^\dag$; $-101.5^\$$ \\
$\chi_{\rm c0}(5P)$ as $X(4700)$ & $-1.1$ & & & & $-54.6^\dag$; $-71.2^\$$ \\
$\chi_{\rm c1}(5P)$ & $-1.0$ & & & & $-76.4^\dag$; $-101.1^\$$ \\
$\chi_{\rm c2}(5P)$ & $-1.2$ & & & & $-78.9^\dag$; $-104.4^\$$ \\
\end{tabular}
\end{ruledtabular}
\caption{As Table \ref{tab:Mass-shifts-4P}, but for $\chi_{\rm c}(5P)$ charmonia. The total self-energies marked by the superscript $\dag$ are the sum of $1P1P$ loop contributions, those marked by $\$$ are the sum of $1P1P$ and also $1P2S$ loop contributions.} 
\label{tab:Mass-shifts-5P}  
\end{table*}

By making use of the above coupled-channel approach, we calculate the relative threshold mass shifts between the $\chi_{\rm c}(4P,5P)$ multiplet members due to a complete set of $(nL, n' L')$ meson-meson loops; see \cite[Sec. 2]{Ferretti:2018tco} and \cite[Sec. IIIC]{Ferretti:2020civ}.
In particular, in the $\chi_{\rm c}(5P)$ case it is easy to identify the relevant set of intermediate states: one has to consider both $1P1P$ meson-meson loops, whose energies range from 4600 MeV [$D_0^*(2300) \bar D_0^*(2300)$] to 5138 MeV [$D_{\rm s2}^*(2573) \bar D_{\rm s2}^*(2573)$], and $1P2S$ loops, whose intermediate state energies span the interval 4864 MeV [$D_0^*(2300) \bar D_0(2550)$] -- 5277 MeV [$D_{\rm s2}^*(2573) \bar D_{\rm s1}^*(2700)$].
In the case of $\chi_{\rm c}(4P)$s, we need to include both $1S2S$ and $1P1P$ loops: this is because the masses of $4P$ charmonia overlap with both $1S2S$ and $1P1P$ intermediate-state energies. We also give results obtained by considering $1S2S$, $1P1P$ and $1S1P$ sets of intermediate states, because the $1S1P$ loops may have an important impact on the properties of the $X(4500)$ as $\chi_{\rm c0}(4P)$.
Furthermore, we neglect charmonium loops, like $\eta_{\rm c} \eta_{\rm c}(2S)$, whose contributions are expected to be very small because of the suppression mechanism of Eq. (\ref{eqn:gamma0-eff}) and \cite[Eq. (12)]{bottomonium} \cite{Ferretti:2020civ}. 

The values of the physical masses, $M_A$, of the $\chi_{\rm c}(4P,5P)$ states should be extracted from the experimental data \cite{Zyla:2020zbs}. 
However, except for the existing $\chi_{\rm c0}(4P,5P)$ candidates, $X(4500)$ and $X(4700)$, nothing is known about the remaining and still unobserved $\chi_{\rm c}(4P,5P)$ states, namely the $h_{\rm c}(4P,5P)$, $\chi_{\rm c1}(4P,5P)$ and $\chi_{\rm c2}(4P,5P)$ resonances.
Therefore, for the physical masses of the previous unobserved states we use the same values as the bare ones; see Table \ref{tab:bare-masses}.
In the case of $\chi_{\rm c0}(4P,5P)$ states, we make the tentative assignments: $\chi_{\rm c0}(4500) \rightarrow \chi_{\rm c0}(4P)$ and $\chi_{\rm c0}(4700) \rightarrow \chi_{\rm c0}(4P)$ or $\chi_{\rm c0}(5P)$.
We thus provide three sets of results for the relative or renormalized threshold corrections, one for each of the previous $\chi_{\rm c0}(4P,5P)$ assignments.
For simplicity, in the present self-energy calculations we do not consider mixing effects between $\left |1^1P_1 \right\rangle$ and $\left|1^3P_1 \right\rangle$ charmed and charmed-strange mesons. Thus, the $\left\langle BC k  \, \ell J \right| T^\dag \left| A \right\rangle$ vertices of Eq. (\ref{eqn:self-a}) are computed under the approximation: $\left |1P_1 \right\rangle \simeq \left |1^1P_1\right\rangle$ and $\left|1P_1' \right\rangle \simeq \left|1^3P_1 \right\rangle$.

\begin{figure}[htbp] 
\centering 
\includegraphics[width=8.5cm]{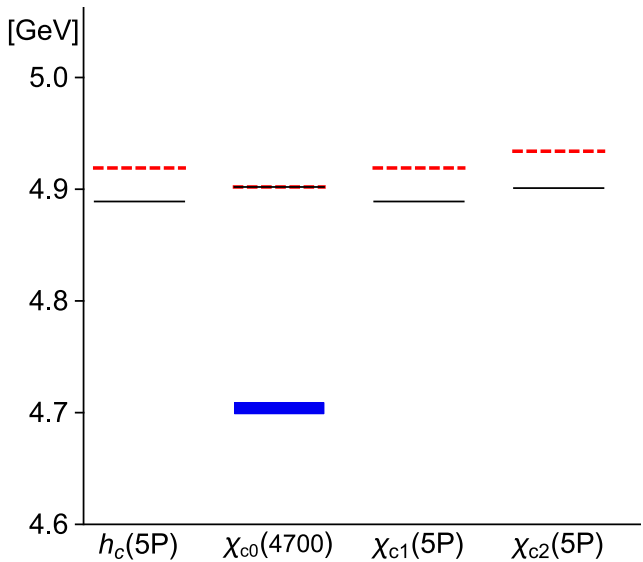}
\caption{Masses of the $\chi_{\rm c}(5P)$ multiplet members with threshold corrections; see Table \ref{tab:ChiC(3P)-splittings}. Here, we consider the assignment $\chi_{\rm c0}(5P) \rightarrow \chi_{\rm c0}(4700)$. 
The blue box stands for the available experimental data \cite{Zyla:2020zbs}, the dashed and continuous lines for the calculated bare and physical masses, respectively. The wide energy gap between the experimental and calculated mass of the $\chi_{\rm c0}(5P)$ state as $\chi_{\rm c0}(4700)$ is a strong indication of the unlikelihood of this assignment.} 
\label{fig:threshold-ChiC(5P)}
\end{figure}

Finally, the self-energy and ``renormalized'' threshold corrections, calculated according to Eqs. (\ref{eqn:self-a}) and (\ref{eqn:new-Ma}), are reported in Tables \ref{tab:ChiC(3P)-splittings}-\ref{tab:Mass-shifts-5P}. 
It is worth noting that: I) the threshold corrections cannot provide an explanation of the discrepancy between the relativized QM value of the $\chi_{\rm c0}(4P)$ mass, 4613 MeV, and the experimental mass of either the $\chi_{\rm c0}(4500)$ or $\chi_{\rm c0}(4700)$ suspected exotics. 
One may attempt to use a different renormalization prescription. For example, in the case of the $\chi_{\rm c0}(4P) \rightarrow \chi_{\rm c0}(4700)$ assignment, one may define the quantity $\tilde \Delta_{\rm th} = \Sigma[\chi_{\rm c2}(4P)]$ rather than $\Delta_{\rm th} = \Sigma[\chi_{\rm c0}(4P)]$ and then plug $\tilde \Delta_{\rm th}$ into Eq. (\ref{eqn:new-Ma}). 
As a result, the calculated physical mass of the $X(4700)$ would be shifted 24 MeV upwards (to 4637 MeV) and would  thus be closer to the experimental value, $4704\pm10^{+14}_{-24}$ MeV \cite{Zyla:2020zbs}. However, the difference between the calculated and experimental masses, 67 MeV, would still be larger than the typical error of a QM calculation, $\mathcal O(30-50)$ MeV; II) something similar happens in the $\chi_{\rm c0}(5P)$ case. Here, the tentative assignment $\chi_{\rm c0}(5P) \rightarrow \chi_{\rm c0}(4700)$ does not work because of the large discrepancy between the calculated and experimental masses of the $\chi_{\rm c0}(5P)$ as $\chi_{\rm c0}(4700)$, namely 4902 MeV and 4704 MeV, respectively; see Fig. \ref{fig:threshold-ChiC(5P)}; III) the renormalized threshold corrections of Table \ref{tab:ChiC(3P)-splittings} are of the order of $20-30$ MeV. The difference between the relativized QM predictions for $\chi_{\rm c0}(4P,5P)$ and the experimental masses of the $\chi_{\rm c0}(4500,4700)$ ranges from $\mathcal O(100)$ MeV in the $\chi_{\rm c0}(4P)$ case to $\mathcal O(200)$ MeV for the $\chi_{\rm c0}(5P)$. 
Because of the wide difference between the data and the QM predictions, the previous threshold corrections do not seem large enough to provide a realistic solution to the mismatch. 
We thus state that the assignment $\chi_{\rm c0}(5P) \rightarrow \chi_{\rm c0}(4700)$ is unacceptable; the tentative assignments $\chi_{\rm c0}(4P) \rightarrow \chi_{\rm c0}(4500)$ or $\chi_{\rm c0}(4700)$ are quite difficult to justify, but cannot be completely excluded.

\section{Conclusion}
We studied the main properties (masses and open-flavor strong decays) of the $4P$ and $5P$ charmonium multiplets.
While there are two candidates for the $\chi_{\rm c0}(4P,5P)$ states, the $X(4500)$ and $X(4700)$ resonances \cite{Zyla:2020zbs,Aaij:2016iza,Aaij:2016nsc}, the properties of the other members of the $\chi_{\rm c}(4P,5P)$ multiplets are still completely unknown.

With this in mind, we first explored the pure charmonium interpretation for these mesons by means of Quark Model (QM) calculations of their open-flavor and radiative decay widths. 
Our QM results, although not conclusive, would suggest the assignments $\chi_{\rm c0}(4500) \rightarrow \chi_{\rm c0}(4P)$ and $\chi_{\rm c0}(4700) \rightarrow \chi_{\rm c0}(5P)$, even if $\chi_{\rm c0}(4700) \rightarrow \chi_{\rm c0}(4P)$ cannot be ruled out completely.

We also discussed the $\chi_{\rm c0}(4500)$ and $\chi_{\rm c0}(4700)$ ``mass problem'', i.e. the incompatibility between the QM predictions for their masses \cite{Godfrey:1985xj} and the experimental data \cite{Zyla:2020zbs,Aaij:2016iza,Aaij:2016nsc}, by making use of a Coupled-Channel Model (CCM) based on the UQM formalism \cite{Ferretti:2018tco,Ferretti:2020civ}.
According to our results for the $\chi_{\rm c0}(4500)$ and $\chi_{\rm c0}(4700)$ masses with threshold/loop corrections, it seems difficult to reconcile the QM predictions with the experimental data, with the possible exception of $\chi_{\rm c0}(4P) \rightarrow \chi_{\rm c0}(4500)$ or $\chi_{\rm c0}(4700)$.

We thus conclude that the $\chi_{\rm c0}(4500)$ and $\chi_{\rm c0}(4700)$ states, which are at the moment excluded from the PDG summary table \cite{Zyla:2020zbs}, are more likely to be described as multiquark states rather than charmonium or charmonium-like ones.

\begin{acknowledgments}
The authors acknowledge financial support from the Academy of Finland, Project no. 320062, and INFN, Italy. 
\end{acknowledgments}

\begin{appendix}

\section{$^3P_0$ pair-creation model}
\label{3P0model}
The $^3P_0$ pair-creation model is an effective model to compute $A \rightarrow BC$ open-flavor strong decays \cite{Micu,LeYaouanc,Roberts:1992}. 
Here, a hadron decay takes place in its rest frame and proceeds via the creation of an additional $q \bar q$ pair. 
The quark-antiquark pair is created with the quantum numbers of the vacuum, i.e. $J^{PC} = 0^{++}$ (see Fig. \ref{fig:diagrammi3P0}), and the decay width can be expressed as \cite{Micu,LeYaouanc,Roberts:1992,Ackleh:1996yt,Ferretti:2013faa,bottomonium}
\begin{equation}
	\label{eqn:3P0-decays-ABC-App}
	\Gamma_{A \rightarrow BC} = \Phi_{A \rightarrow BC}(k_0) \sum_{\ell} 
	\left| \left\langle BC k_0  \, \ell J \right| T^\dag \left| A \right\rangle \right|^2 \mbox{ }.
\end{equation}
The final state is characterized by the relative orbital angular momentum $\ell$ between $B$ and $C$ and a total angular momentum ${\bf J} = {\bf J}_B + {\bf J}_C + {\bm \ell}$.
One usually assumes harmonic oscillator wave functions for the parent, $A$, and daughter, $B$ and $C$, hadrons, depending on a h.o. parameter, $\alpha_{\rm ho}$. In the meson case, one has
\begin{eqnarray}
	\label{eqn:3DHO-Pc}
	\Phi_{n_j l_j m_j} ({\bf q}_j) & = & \mathcal{N}_{n_j l_j}(\alpha_j) 
	L_{n_j}^{l_j+1/2}(\alpha_j^{-2} q_j^2) \mbox{ } e^{-\frac{1}{2} q_j^2/\alpha_j^2} \nonumber\\ 
	& & \mathcal Y_{l_j m_j} ({\bf q}_j) \mbox{ },
\end{eqnarray}
where the index $j = A, B$ or $C$ distinguishes among parent and daughter mesons, $\alpha_j = \alpha_{\rm ho}$, $n_j$ is the radial quantum number, $L_{n_j}^{l_j+1/2}(\alpha_j^{-2} q_j^2)$ a Laguerre polynomial and $\mathcal Y_{l_j m_j} ({\bf q}_j)$ a solid spherical harmonic \cite{Edmonds}; 
\begin{equation}
	\label{eqn:3DHO-rc}
	\mathcal{N}_{n_j l_j}(\alpha_j) = \sqrt{\frac{2 n_j !}{\Gamma(n_j+l_j+3/2)}} \mbox{ } \alpha_j^{-l_j-\frac{3}{2}} 
\end{equation}
is the normalization factor of the h.o. wave function of Eq. (\ref{eqn:3DHO-Pc}).
The coefficient $\Phi_{A \rightarrow BC}(k_0)$ in Eq. (\ref{eqn:3P0-decays-ABC-App}) is the phase-space factor (PSF) for the decay.
Several prescriptions for $\Phi_{A \rightarrow BC}(k_0)$ are possible \cite{Capstick:2000qj}, including the non-relativistic one,
\begin{equation}
	\label{eqn:nonrel-PSF}
	\Phi_{A \rightarrow BC}(k_0) = 2 \pi k_0 \frac{M_B M_C}{M_A}  \mbox{ },
\end{equation}
depending on the relative momentum $k_0$ between $B$ and $C$ and on the masses of the parent, $M_A$, and daughter hadrons, $M_B$ and $M_C$. The second option is the standard relativistic form,
\begin{equation}
	\label{eqn:rel-PSF}
	\Phi_{A \rightarrow BC}(k_0) = 2 \pi k_0 \frac{E_B(k_0) E_C(k_0)}{M_A}  \mbox{ },
\end{equation}
where $E_B = \sqrt{M_B^2 + k_0^2}$ and $E_C = \sqrt{M_C^2 + k_0^2}$ are the energies of the daughter hadrons.
The third possibility is to use an effective PSF \cite{Capstick:1992th,Kokoski:1985is},
\begin{equation}
	\label{eqn:eff-PSF}
	\Phi_{A \rightarrow BC}(k_0) = 2 \pi k_0 \frac{\tilde M_B \tilde M_C}{M_A}  \mbox{ },
\end{equation}
where $\tilde M_B$ and $\tilde M_C$ are the effective masses of the daughter hadrons, evaluated by means of a spin-independent interaction.
Our choice here is to use the PSF of Eq. (\ref{eqn:rel-PSF}).
However, it is worth to note that in the case of heavy baryons and mesons, whose internal dynamics is almost non-relativistic and the hyperfine interactions are relatively small, the three types of phase-space factors are expected to provide very similar results.

The transition operator of the $^{3}P_0$ model is given by \cite{Micu,LeYaouanc,Roberts:1992,Strong2015,Bonnaz:1999zj}:
\begin{equation}
	\label{3p0}
	\begin{array}{rcl}
	T^{\dagger} & = & -3 \gamma_0 \, \int d {\bf p}_3 \, d {\bf p}_4 \, \delta({\bf p}_3 + {\bf p}_4) \, C_{34} \, 
	F_{34} \, V({\bf p}_3 - {\bf p}_4) \\
	& \times & \left[ \chi_{34} \, \times \, {\cal Y}_{1}({\bf p}_3 - {\bf p}_4) \right]^{(0)}_0 \, 
	b_3^{\dagger}({\bf p}_3) \, d_4^{\dagger}({\bf p}_4)    \mbox{ }.
	\end{array}
\end{equation}
Here, $\gamma_0$ is the pair-creation strength, whose value is fitted to the reproduction of the experimental strong decay widths \cite{Zyla:2020zbs}; $b_3^{\dagger}({\bf p}_3)$ and $d_4^{\dagger}({\bf p}_4)$ are the creation operators for a quark and an antiquark with momenta ${\bf p}_3$ and ${\bf p}_4$, respectively; see Fig. \ref{fig:diagrammi3P0}.
The created $q \bar q$ pair is characterized by a color-singlet wave function, $C_{34}$, a flavor-singlet wave function, $F_{34}$, a spin-triplet wave function $\chi_{34}$, and a solid spherical harmonic ${\cal Y}_{1}({\bf p}_3 - {\bf p}_4)$, because the quark and antiquark are in a relative $P$-wave. 

$V({\bf p}_3 - {\bf p}_4)$ in Eq. (\ref{3p0}) is the pair-creation vertex (PCV). In the original formulation of the $^3P_0$ model \cite{Micu,LeYaouanc}, one has $V({\bf p}_3 - {\bf p}_4) = 1$. 
Possible refinements of the PCV were discussed e.g. in Refs. \cite{Strong2015,Bonnaz:1999zj}.
In the present calculations, we consider the PCV or quark form factor (QFF) \cite{bottomonium,Geiger:1989yc,Bijker:2009up,Strong2015}
\begin{equation}
	V({\bf p}_3 - {\bf p}_4) = e^{- ({\bf p}_3 - {\bf p}_4)^2/(6\alpha_{\rm d}^2)}  \mbox{ },
\end{equation}	
where $\alpha_{\rm d}$ is the QFF parameter. The introduction of this particular PCV is motivated by the request that the $q \bar q$ pair of created quarks has an effective size.

In addition to that, another modification of the original $^3P_0$ model is worth to be taken into account.
It consists in the substitution of the constant pair creation strength $\gamma_0$ of Eq. (\ref{3p0}) with an effective, flavor-dependent, pair-creation strength, 
\begin{equation}
	\label{eqn:gamma0-eff}
	\gamma_0^{\rm eff} = \frac{m_{\rm u,d}}{m_{\rm i}} \mbox{ } \gamma_0  \mbox{ }.
\end{equation}
Its purpose is to suppress unphysical heavy quark pair-creation \cite{Kalashnikova:2005ui,Ferretti:2013faa,bottomonium,Strong2015}. The mechanism consists in multiplying $\gamma_0$ by a reduction factor $m_{\rm u,d}/m_{\rm i}$, with $i = u, d, s$ or $c$; see Table \ref{tab:3P0parameters}. For example, the creation of $s \bar s$ pairs is suppressed with respect to $u \bar u$ ($d \bar d$) by a factor of $(m_{\rm u,d}/m_{\rm s})^2 = 0.36$, while that of $c \bar c$ by one of $(m_{\rm u,d}/m_{\rm c})^2 = 0.05$.
In the case of $m_{\rm i} = m_{\rm u}$ or $m_{\rm d}$, $\gamma_0^{\rm eff} = \gamma_0$ and no suppression occurs.
It is worth to note that this particular choice for the pair-creation strength breaks the $SU(3)$ [$SU(4)$] flavor symmetry and its effect cannot be re-absorbed in a redefinition of the model parameters or in a different choice of the $^3P_0$ model vertex factor \cite{Strong2015}.

Finally, the $\left\langle BC k_0  \, \ell J \right| T^\dag \left| A \right\rangle$ $^3P_0$ amplitudes of Eq. (\ref{eqn:3P0-decays-ABC-App}) can be calculated analytically by means of the formalism discussed in Ref. \cite{Roberts:1992}. For the open-flavor strong decays of baryons, see also Ref. \cite{Strong2015}.

\end{appendix}

\end{document}